\documentclass
[twocolumn,showpacs,preprintnumbers,amsmath,amssymb,aps,pre,superscriptaddress,floatfix,reprint]
{revtex4}
\usepackage{graphicx}
\usepackage{dcolumn}
\usepackage{bm}
\usepackage{here}
\usepackage{color}
\usepackage{braket}
\begin{document}

\newcommand{\fracd}[2]{\frac{\displaystyle #1}{\displaystyle #2}}
\newcommand{\red}[1]{\textcolor{red}{#1}}
\newcommand{\blue}[1]{\textcolor{blue}{#1}}
\newcommand{\green}[1]{\textcolor{green}{#1}}

\title{Critical Phenomena of Dynamical Delocalization in Quantum Maps:
standard map and Anderson map}
\author{Hiroaki S. Yamada}
\affiliation{Yamada Physics Research Laboratory,
Aoyama 5-7-14-205, Niigata 950-2002, Japan}
\author{Kensuke S. Ikeda}
\affiliation{College of Science and Engineering, Ritsumeikan University
Noji-higashi 1-1-1, Kusatsu 525-8577, Japan}

\date{\today}
\begin{abstract}
Following the paper exploring the Anderson localization of monochromatically 
perturbed kicked quantum maps [Phys.Rev. E{\bf 97},012210], 
the delocalization-localization transition phenomena in polychromatically perturbed quantum maps (QM) is investigated 
focusing particularly on the dependency of critical phenomena 
upon the number $M$ of the harmonic perturbations, where $M+1=d$ corresponds to the spatial 
dimension of the ordinary disordered lattice.
The standard map and the Anderson map are treated and compared.
As the basis of analysis, we apply the self-consistent theory (SCT) of the localization, 
taking a plausible hypothesis on the mean-free-path parameter 
which worked successfully 
in the analyses of the monochromatically perturbed QMs. 
We compare in detail the numerical results with the predictions of the SCT, 
by largely increasing $M$. The numerically obtained index of critical subdiffusion 
$t^\alpha$~($t$:time) agrees well with the prediction of one-parameter scaling 
theory $\alpha=2/(M+1)$, 
but the numerically obtained critical exponent of localization length 
significantly deviates from the SCT prediction. Deviation from the SCT prediction 
is drastic for the critical  perturbation strength of the transition: if $M$ is 
fixed the SCT presents plausible prediction for the 
parameter dependence of the critical value, but its value is $1/(M-1)$times 
smaller than the SCT prediction, which implies existence of a strong cooperativity 
of the harmonic perturbations with the main mode. 
\end{abstract}



\maketitle


\newcommand{\vc}[1]{\mbox{\boldmath $#1$}}
\newcommand{\del}{\partial}

\def\ni{\noindent}
\def\nn{\nonumber}
\def\bH{\begin{Huge}}
\def\eH{\end{Huge}}
\def\bL{\begin{Large}}
\def\eL{\end{Large}}
\def\bl{\begin{large}}
\def\el{\end{large}}
\def\beq{\begin{eqnarray}}
\def\eeq{\end{eqnarray}}

\def\eps{\epsilon}
\def\th{\theta}
\def\del{\delta}
\def\omg{\omega}

\def\e{{\rm e}}
\def\exp{{\rm exp}}
\def\arg{{\rm arg}}
\def\Im{{\rm Im}}
\def\Re{{\rm Re}}

\def\sup{\supset}
\def\sub{\subset}
\def\a{\cap}
\def\u{\cup}
\def\bks{\backslash}

\def\ovl{\overline}
\def\unl{\underline}

\def\rar{\rightarrow}
\def\Rar{\Rightarrow}
\def\lar{\leftarrow}
\def\Lar{\Leftarrow}
\def\bar{\leftrightarrow}
\def\Bar{\Leftrightarrow}

\def\pr{\partial}

\def\>{\rangle} 
\def\<{\langle} 
\def\RR {\rangle\!\rangle} 
\def\LL {\langle\!\langle} 
\def\const{{\rm const.}}

\def\e{{\rm e}}

\def\Bstar{\bL $\star$ \eL}

\def\etath{\eta_{th}}
\def\irrev{{\mathcal R}}
\def\e{{\rm e}}
\def\noise{n}
\def\hatp{\hat{p}}
\def\hatq{\hat{q}}
\def\hatU{\hat{U}}
\def\hatJ{\hat{J}}
\def\hatphi{\hat{\phi}}
\def\hatx{\hat{x}}
\def\haty{\hat{y}}
\def\hatF{\hat{F}}
\def\hatG{\hat{G}}

\def\hatA{\hat{A}}
\def\hatB{\hat{B}}
\def\hatC{\hat{C}}
\def\hatJ{\hat{J}}
\def\hatI{\hat{I}}
\def\hatP{\hat{P}}
\def\hatQ{\hat{Q}}
\def\hatU{\hat{U}}
\def\hatW{\hat{W}}
\def\hatX{\hat{X}}
\def\hatY{\hat{Y}}
\def\hatV{\hat{V}}
\def\hatt{\hat{t}}
\def\hatw{\hat{w}}

\def\hatp{\hat{p}}
\def\hatq{\hat{q}}
\def\hatU{\hat{U}}
\def\hatn{\hat{n}}

\def\hatphi{\hat{\phi}}
\def\hattheta{\hat{\theta}}

\def\iset{\mathcal{I}}
\def\fset{\mathcal{F}}
\def\pr{\partial}
\def\traj{\ell}
\def\eps{\epsilon}
\def\U{\hat{U}}

\def\U{U_{\rm cls}}
\def\P{P_{{\rm cls},\eta}}
\def\traj{\ell}
\def\cc{\cdot}

\def\DZ{D^{(0)}}
\def\Dcls{D_{\rm cls}}

\newcommand{\relmiddle}[1]{\mathrel{}\middle#1\mathrel{}}

\section{Introduction} 
It is a basic nature of
the freely propagating quantum particle that it localizes 
by inserting random impurities \cite{anderson58,ishii73}
and its normal conduction,
which is an irreversible quantum Brownian motion, is realized 
after destroying the localization by some additional operations.    
The ordinary way to free from localization is to increase the spatial
dimension of the system and weaken the randomness. 
An another way is to introduce dynamical perturbations such as 
harmonic vibrations due to the lattice vibration. The destruction
of localization by the latter way is called dynamical delocalization.
The purpose of the present paper is to elucidate the critical
phenomena of the dynamical localization-delocalization
transition (LDT) numerically and theoretically, 
following previous papers \cite{yamada15} and \cite{yamada18}.
(We referred to them as [I] and [II], respectively, in the text.)
%
%
Recently the localization and delocalization of wavepacket propagation
has been investigated experimentally and theoretically.
In particular, the quantum standard map (SM) systems, which 
theoretically shown to exhibit dynamical localization \cite{casati79}, 
has been studied extensively. If SM is coupled with dynamical harmonic 
perturbations composed of $M$ incommensurate frequencies, it
can formally be transformed into a $d(=M+1)-$dimensional lattice
system with quasi-periodic potential 
\cite{casati89,borgonovi97,chabe08,wang09,lemarie10,tian11}.
Then it can be expected
that the harmonically perturbed SM will undergo a Anderson
transition of the $d(=M+1)-$dimensional random quantum lattice.

Indeed,  Lopez {\it et al} implemented the perturbed SM as a cold atom 
on the optical lattice, and succeeded in observing the Anderson
transition \cite{lopez12,lopez13}. They obtained the critical diffusion exponents
and the critical localization exponents experimentally, which agreed 
with numerical and theoretical results for $M=2$.   
They also observed an exponentially  exteded localization for $M=1$ \cite{manai15}.


We can then expect that even the localization phenomenon on
low-dimensional disordered quantum lattice can be also
delocalized by applying harmonic perturbations with finite
number of incommensurate frequency components \cite{yamada98,yamada99}. 
The increment
of the number $M$ of the frequencies will make the delocalization
easier, thereby realizing the onset of diffusion which is
a typical irreversible motion simulating the normal conduction of electron.
To examine the above conjecture, we proposed a quantum map defined
on a disordered lattice, which we call the Anderson map (AM) \cite{yamada04}. 
It evolves in a discretized time and become the one-dimensional disordered system 
in a continuous time limit.


The SM of $M=1$ corresponds to the asymmetric two-dimensional
disordered system, and the localization length is exponentially
enhanced but the LDT does not occur,
which has been confirmed experimentally and numerically \cite{manai15}. 
In the previous paper [II]\cite{yamada18} we also numerically and theoretically studied the
localization characteristics of AM of $M=1$ in comparison
with that of the SM of $M=1$, and all the numerical results were 
well explained in terms of the self-consistent theory (SCT) 
of the localization \cite{wolfle10}.
 The AM of $M=1$
 has a paradoxical character that the localization length increases
as the disorder strength $W$ of potential  exceeds a threshold value
$W^*$, which was successfully predicted by the SCT.

On the other hand, we presented a preliminary paper [I] in which we showed
that the AM with $M\geq 2$ undergoes the LDT
as is the case of the SM with $M\geq 2$ and further the results based
upon the one-parameter scaling hypothesis can explain the critical
diffusion exponent for a wide range of $M$ \cite{yamada15}. 

The present paper provides a complete numerical and theoretical analysis of the 
localization-delocalization characteristics of AM in comparison with SM for a 
wide range of control parameters, particularly, with changing $M$ largely.

In Sect.\ref{sec:model}, we introduce polychromatically perturbed quantum 
standard map (SM) and Anderson map (AM).
First, in Sect.\ref{sec:subdiffusion} we begin with reviewing the results  
reported by the papers [I] and paper [II] about
the $M$ dependency of the critical subdiffusion exponent 
and the critical localization exponent, including some new results.
We are particularly interested in the dependencies of the 
critical perturbation strength of the harmonic perturbation (we denote
it by $\eps_c$ hereafter) on the control parameters of the system 
and the predictability of the SCT for them.
We show in Sect.\ref{sec:epsc} the theoretical prediction based on SCT
for critical perturbation strength $\eps_c$  of the LDT for SM and AM 
and compare them with the numerical results. Except for $M$, the SCT
successfully predicts the dependency of $\eps_c$ upon the control parameters.
However, the SCT fails to predict the $M-$dependence.
Numerically, it turns out that $\eps_c \simeq 1/(M-1)$ for both AM and
SM, but the SCT predicts that it is a constant.
In Sect.\ref{sec:summary}, we summarize and discuss the result.
The derivation of some equations and some details of the numerically decided 
critical exponent of the localization are given in appendixes.


\section{Models and their dynamics} 
\label{sec:model}
We consider dynamics of the following quantum map systems 
represented by the Hamiltonian,
\beq
H_{tot}(\hatp,\hatq,t) =T(\hatp) + V(\hatq,\{\omega_jt\}) \delta_t, 
\label{eq:Htot}
\eeq
where $\delta_t=\sum_{k=-\infty}^{\infty}\delta(t-k\Delta)$.
In this paper we set the period of the kicks $\Delta=1$.
$T(\hatp)$ is the kinetic energy term, and the potential energy term 
$V(\hatq,t) $ including time dependent perturbation $f(t)$ is given as,
\beq
V(\hatq,\{\omega_jt\}) & =& V(\hatq) [1+ f(\{\omega_jt\})]  \\
&=& V(\hatq) \left[1+ \frac{\eps}{\sqrt{M}} \sum_j^M \cos(\omega_j t)\right] ,
\label{eq:perterbation}
\eeq
where $M$ and $\eps$ are number of the frequency component and 
the strength of the perturbation, respectively.
Note that the strength of the perturbation is divided by $\sqrt{M}$
so as to make the total power of the long-time average 
independent of $M$, i.e. $\overline{f(\{\omega_it\})^2}=\eps^2/2$, and 
the frequencies $\{ \omega_j\}(j=1,...,M)$ are taken 
as mutually  incommensurate number of $O(1)$.
Here $\hatp$ and $\hatq$ are momentum and position
operators, respectively. 

In the present paper, we use the standard map (SM), which is
given by,
\beq
T(\hatp)= \frac{p^2}{2},~~V(\hatq)=v(\hat q).
\eeq
In addition, we deal with Anderson map (AM), which is given by,
\beq
T(\hatp)= 2\cos(\hatp/\hbar),~~V(\hatq)=Wv(\hatq),
\eeq
where 
\beq
  v(\hatq)=
\begin{cases}
            K \cos \hatq  &  {\rm (for~SM)} \\ 
           \sum_{n \in {\Bbb Z}}\delta(q-n) v_n|n\>\<n|  & {\rm (for~AM)}.
\end{cases}
\eeq
In the case of SM the global propagation occurs in the momentum space $p$ 
spanned by the momentum eigenstates $|p\>=|P\hbar \>~~(P\in {\Bbb Z})$,
being transferred by the potential operator $v(\hatq)$. On the other hand, in the case of 
AM $v(\hatq)$ plays the role of the on-site potential operator 
taking random value $v_n$ uniformly distributed 
over the range $[-1, 1]$, and $W$ denotes the disorder strength.
The global
propagation occurs in the position space $q$, which are 
spanned by the position eigenstates $|n\>~(n\in {\Bbb Z})$ \cite{yamada10}.
The AM is a quantum map with discretized time but it approaches to the time-continuous
Anderson model defined on the random lattice for $W \ll 1$.

We can regard the harmonic perturbations as the dynamical degrees of freedom. To show this
we introduce the classically canonical action-angle operators 
$(\hatJ_j=-i\hbar \frac{\pr_j}{\pr \phi_j}, \phi_j)$
representing the harmonic perturbation as a linear mode (we call the ``harmonic mode'' hereafter)
and extend the Hamiltonian (\ref{eq:Htot}) so as to include the harmonic modes,
\beq
&& H_{aut}(\hatp,\hatq,\{\hatJ_j\},\{\hatphi_j\})  \nn \\
&&
  ~~~~~~=T(\hatp) + V(\hatq,\hatphi,\{\hatphi_j\}) \delta_t+\sum_{j=1}^M \omega_j\hatJ_j, 
\label{eq:Haut}
\eeq
where 
\beq
 V(\hatq,\{\hatphi_j\}) &=& V(\hatq) [1+ f(\{\hatphi_j\})],  \\
\nn  &=& V(\hatq) \left[1+ \frac{\eps}{\sqrt{M}} \sum_j^M \cos \phi_j \right].
\eeq
One can easily check that by Malyland transform
the eigenvalue problem of the quantum map system interacting with 
$M$-harmonic modes can be transformed into $d(=M+1)-$dimensional
lattice problem with quasi-periodic and/or 
random on-site potentials \cite{fishman82,yamada18}. 
(See appendix \ref{app:Maryland}.)
In this view, to increase the number of the harmonic modes is to
increase the dimension of the system, which enables 
the LDT.

From the dynamical point of view, the harmonic modes perturbs the main mode to  cause 
the diffusive motion and induce the LDT. On the other hand, by the backaction of the 
perturbation to the main mode, the harmonic mode is excited to propagate along the ladder of action eigenstates
satisfying $\hatJ_j|m_j\>=m_j\hbar|m_j\>~(m_j \in {\Bbb Z}$).
Let $\haty_j=\hatJ_j/\hbar=\sum_{m_i \in {\Bbb Z}} m_i |m_i\>\<m_i|$ 
be the operator indicating
the excitation number in the action space, then the Heisenberg equation of 
motion $d\haty_j/dt\hbar=(i/\hbar^2)[H_{aut},\hatJ_j]=-1/\hbar \pr V(\hatq,\{\phi_j\})/\pr\phi_j$ gives
the step-by-step evolution rule for the Heisenberg operators: 
\beq
\label{eq:propcolor}
   \haty_j(t)-\haty_j(0) =\frac{\eps}{\sqrt{M}} \sum_{s=0}^{t}C_j\hatG(s)\sin(\omega_js+\phi_{j0}),
\eeq
where $\phi_{j0}$ is the initial phase. Here, 
\beq 
\label{eq:propcolorforce-1}
&&    \hatG(t)=\frac{1}{\hbar}v(\hatq(t))
\eeq
and
\beq
\label{eq:propcolorforce-2}
C_j= 
   \begin{cases}
       1~~~{\rm~~(for~SM)},    \\
       W~~{\rm~~(for~AM)}.
   \end{cases}
\eeq
The potential $v(\hatq(t))$ works as a force inducing a propagation along the action ladder.

To treat the transport in the main mode of SM and AM in a unified manner, we define
the excitation number operator in the momentum space $\hatx=\hatp/\hbar=\sum_{P} P|P\hbar \>\<P\hbar|~~(P\in {\Bbb Z})$
for SM and in the real space $\hatx=\sum_nn|n\>\<n|~(n\in {\Bbb Z})$ for AM, where $|P\hbar\>$ and $|n\>$
are the momentum and the real position eigenstates, respectively. Then the step-by-step evolution
rule for the Heisenberg operator is
\beq
\label{eq:propmain}
    \hatx(t)-\hatx(0) =  \sum_{s=0}^{t}\hatF(s),
\eeq
where the force $\hatF$ is 
\beq
\label{eq:propmainforce}
\hatF(t) = 
   \begin{cases}
\frac{K}{\hbar}\sin\hatq(t)~~& {\rm (for~SM)}     \\
     -\frac{2}{\hbar}\sin(\hatp(t)/\hbar)                      ~~& {\rm (for~AM)} .
   \end{cases}
\eeq
In the next section, 
with the basic formal representations presented above, we first discuss the localization
of unperturbed SM and AM and further the transition to the delocalized states.

\section{Critical subdiffusion of LDT in the polychromatically perturbed 
quantum maps}
\label{sec:subdiffusion}
In this section we show the results related to the critical
subdiffusion which is a remarkable feature of the critical state of the 
LDT, by organizing the known results 
reported in the previous papers \cite{yamada18,yamada15} and the new ones.

\subsection{Localization in the unperturbed and monochromatically perturbed 
quantum maps ($M=0,1$)}
We use an initial quantum state $|\Psi(t=0)\>$ and the $x-$representation 
$\<x|\Psi(t=0)\>=\delta_{x,N/2}$ and characterize quantitatively the spread 
of the wavepacket by the mean square displacement (MSD),
\beq
\label{eq:m2org}
  m_2(t)&=&\<\Psi(t=0)|(\hatx(t)-\hatx(0))^2|\Psi(t=0)\> \nn \\
&\equiv &\<(\hatx(t)-\hatx(0))^2\>, 
\eeq
where $\hatx$ is $\hatp$ for SM and $\hatx$ is $\hatq$ for AM, respectively.
Using Eq.(\ref{eq:propmain}), it immediately follows that
\beq
\label{eq:m2}
m_2(t)=\sum_{s\leq t}\DZ_0(s:t),
\eeq
where
\beq
\DZ_0(s:t)= \sum_{s^{'}=s}^{t}\<\hatF(s)\hatF(s^{'})\>+c.c.
\eeq

In the unperturbed 1D quantum maps with $\eps=0$, the time-dependent diffusion constant
$\DZ_0(s:t)$, which converges to a positive finite value $\DZ_0$ if $s$ is small and 
$t\to \infty$, finally goes to zero as $s$ increases, and thus 
$m_2(t)$ given by Eq.(\ref{eq:m2}) saturates and the wavepacket become 
localized in the limit $t\to\infty$. Let the localization length and
the time scale beyond which the diffusion terminates be $\ell_0$ and
$t_0$, respectively, then Eq.(\ref{eq:m2}) gives $\ell_0^2=m_2(\infty)\sim
\DZ_0t_0$, where $\DZ_0:= \DZ_0(0:\infty)$ is the initial stage diffusion 
constant which converges to a positive finite value.

In the localized phase, in the spatial region of localization length $\ell_0$ all the localized
eigenfunctions of number $\ell_0$ supported by the region undergo very strong level repulsion.
The interval between the nearest neighbouring eigenangles should be $\sim 1/\ell_0$, 
which means that its inverse ($\sim \ell_0$) characterizes the localization time $t_0$. 
Then the relation  means that
\beq
   \ell_0^2 \sim \DZ_0 \ell_0   \label{eq:m2-loc},
\eeq
and therefore
\beq
  \DZ_0\sim \ell_0 \sim t_0  \label{eq:m2-loc-D}
\eeq
the localization length as well as the localization time are decided by 
the diffusion constant.
One can confirm that the SCT discussed later 
also supports the above relation if it is applied to the isolated 
(i.e., $\eps=0$) one-dimensional system.
In the case of isolated SM,  $\DZ_0$ equals to the classical chaotic 
diffusion constant \cite{casati84}:
\beq
\label{eq:m2-loc-D-SM}
   \DZ_0 \sim \ell_0 \sim D_{cls}/\hbar^2~~\to K^2/\hbar^2~(K^2\gg1) 
\eeq

On the other hand, in the case of the isolated AM, the well-known result $\ell_0\sim 1/W^2$
for the continuous-time Anderson model holds \cite{lifshiz88} . 
However, this result holds
correct only for $W$ less than the characteristic value decided by 
\beq
\label{eq:m2-Wstar-AM}
   W^* \sim 2\pi \hbar,
\eeq
beyond which $\ell_0$ terminates to decrease and approaches to 
a constant $\sim 1/W^{*2}$ \cite{yamada18}.
This is a remarkable feature of the AM different from the continuous-time Anderson model. 
Then, we have
\beq
\label{eq:m2-loc-D-AM}
 \DZ_0 \sim \ell_0 \sim 
  \begin{cases}
    & 1/W^{2}~~~~(W\ll W^*) \\
    & 1/W^{*2}~~~(W\gg W^*).
  \end{cases}
\eeq
A basic hypothesis assumed here is that
the temporal localization process of isolated system starts with 
a transient diffusion process
with the diffusion constant $\DZ_0$. As will be discussed
later this hypothesis does not work in a certain case of
AM, but we first use this hypothesis in the next section. 
As is shown in the appendix \ref{app:Maryland}, the eigenvalue problem 
of our systems, which are represented as $M+1$ degrees of freedom system
in the extended scheme of Eq.(\ref{eq:Haut}), is formally transformed
into $d(=M+1)-$dimensional lattice problem with quasi-periodic and/or 
random on-site potentials by the so-called Maryland transform. 
As was demonstrated in
the paper [I] the delocalization transition do not occur for $M=1$,
i.e., for the effective dimension $d=2$, although the localization
length grows exponentially as $\ell_0 \propto \e^{{\rm const.} \eps}$.
We thus consider the case $M(=d-1) \geq 2$, for which the 
LDT may take place 
according to the ordinary scenario of Anderson transition.

\subsection{$M-$dependence of subdiffusion in SM and AM  ($M \geq 2$)}
As partially shown in the paper [II], 
the perturbation strength $\eps$ exceeds the critical value $\eps_c$ 
the LDT occurs if $M\geq 2$.  

In the LDT an anomalous diffusion 
\beq
m_2 \sim t^\alpha (0<\alpha<1).
\eeq
with the characteristic exponent $\alpha$ is observed at the critical perturbation strength
 $\eps=\eps_c$.


The presence of subdiffusion is confirmed in the preliminary report [II], 
and a more detailed study of the critical subdiffusion 
for control parameters covering much wider regime is executed.
It is convenient to define the scaled MSD $\Lambda(t)$ divided by
the critical subdiffusive factor in order to investigate the critical 
behavior close to LDT:
\beq
  \Lambda(t)=\frac{m_2(t)}{t^\alpha}.
\eeq
This scaled MSD is also used in finite-time scaling to determine 
the critical exponent of LDT.
(See appendix \ref{app:critical-exponent}.)



\begin{figure}[htbp]
\begin{center}
\includegraphics[width=8cm]{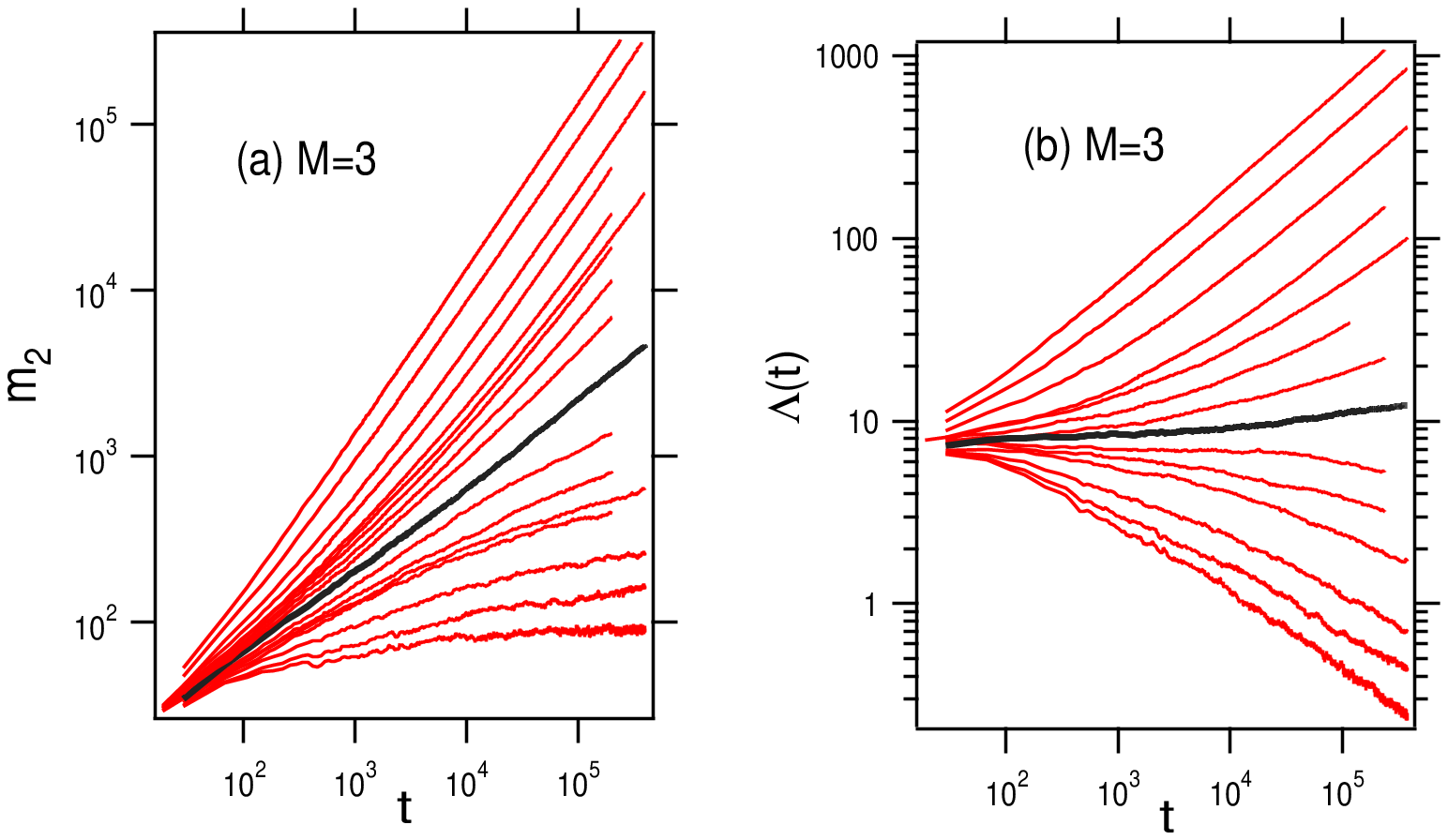}
\hspace{2mm}
*\hspace{8mm}
\includegraphics[width=8cm]{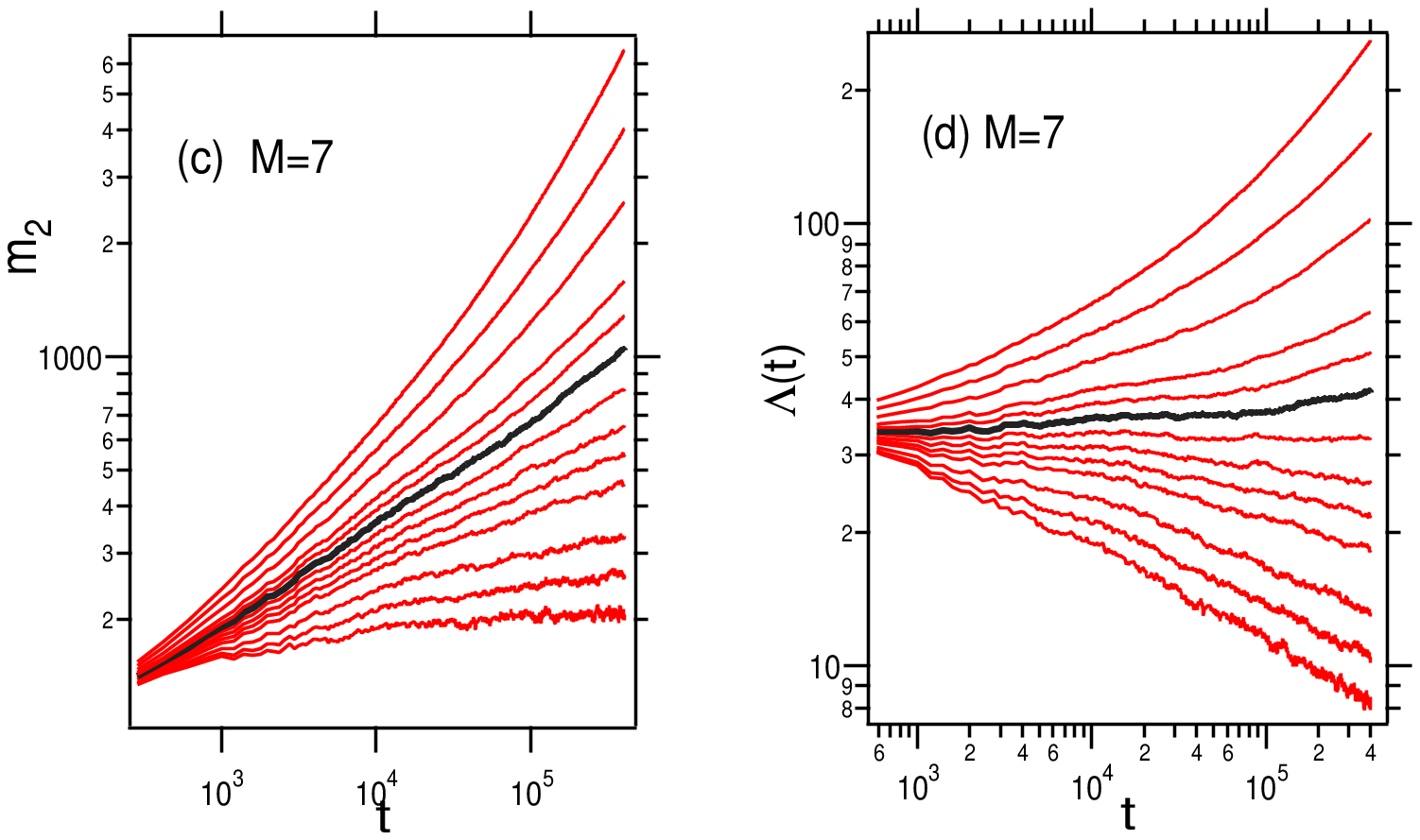}
\caption{(Color online)
The double-logarithmic plots of (a)$m_2(t)$ 
and (b)the scaled $\Lambda(\eps,t)$ 
as a function of time
for different values of the perturbation strength $\eps$, 
where the diffusion exponent $\alpha$ is determined by 
the least-square-fit for the $m_2(t)$ with the critical case, 
in the polychromatically perturbed SM of $M=3$ with $K=3.1$,
$\hbar=2\pi\times 311/2^{13}(\equiv \hbar_0)$.
(c)The same $m_2(t)$ and (d)the scaled $\Lambda(\eps,t)$ 
in the polychromatically perturbed SM of $M=7$.
In the case of $M=3$, 
$\eps_c^{SM} \simeq 0.0081$, $m_2 \sim t^\alpha$ with $\alpha \simeq 0.46$.
In the case of $M=7$, 
$\eps_c^{SM} \simeq 0.012$, $m_2 \sim t^\alpha$ with $\alpha \simeq 0.25$.
The data near the critical value $\eps_c$ are shown by bold black lines.
In the following we representation 
$\hbar=\hbar_0, 2\hbar_0, 3\hbar_0,...$ 
as an unit $\hbar_0=2\pi\times 311/2^{13}\simeq0.24$.
}
\label{fig:SM-MSD}
\includegraphics[width=9cm]{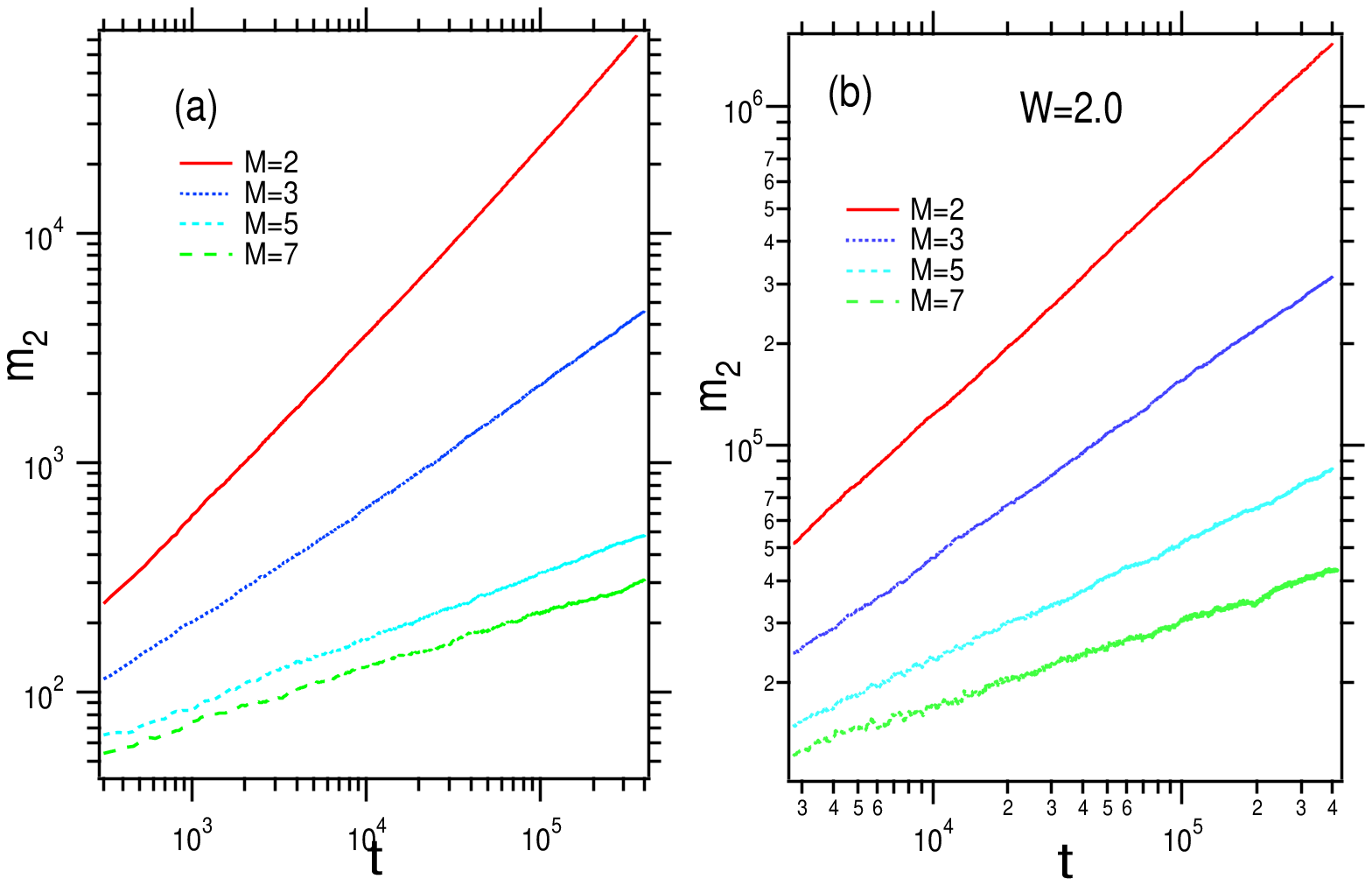}
\caption{(Color online)
The double-logarithmic plots of $m_2(t)$ 
as a function of time near the critical pints $\eps_c$ in
(a) the polychromatically perturbed SM ($M=2, 3, 5, 7$) with $K=3.1$,
$\hbar=\hbar_0$, 
and 
(b) AM ($M=2, 3, 5, 7$) with $W=2.0$.
In the perturbed SM and AM, the system and ensemble sizes are 
$N=2^{15} \sim 2^{17}$ and $10 \sim 100$, respectively,  
throughout this paper.
}
\label{fig:typical-msd}
\end{center}
\end{figure}

We first show the case of SM. 
Figure \ref{fig:SM-MSD}(a) and (c) show the time-dependence of MSD $m_2(t)$
in the cases of $M=3$ and $M=7$, respectively, for various values of $\eps$ 
increasing across the critical value $\eps_c$.
Figure \ref{fig:SM-MSD} (b) and (d) show the scaled MSD $\Lambda(t)$ corresponding
to  (a) and (c).
It can be seen that a transition from the localized state to the delocalized state 
occurs going through a stable subdiffusion state as $\eps$ increases.
The scaled MSD $\Lambda(t)$ also shows a very characteristic holding-fan-pattern whose 
behavior leads to a remarkable scaled behavior with respect to 
the critical parameter $|\eps-\eps_c|$.

Figure \ref{fig:typical-msd}(a) shows the critical subdiffusions at the critical point $\eps=\eps_c$ 
when the color number $M$ is changed. It is evident that the diffusion index $\alpha$ at 
the critical point $\eps_c$ decreases as $M$ increases, and the numerical results tell
that it can be approximated very well by the rule
\beq
\alpha \simeq \frac{2}{M+1}, 
\label{eq:alpha}
\eeq
regardless of the values of the control parameters such as $K$ and $\hbar$.
The result is also consistent with the well-known guess 
based upon the one-parameter scaling theory (OPST) of the localization, 
which are summarized in appendix \ref{app:OPST}.
The critical value $\eps_c$ decreases with $M$ as well as $\alpha$, 
which will be discussed in detail in next section.

\begin{figure}[htbp]
\begin{center}
\includegraphics[width=8cm]{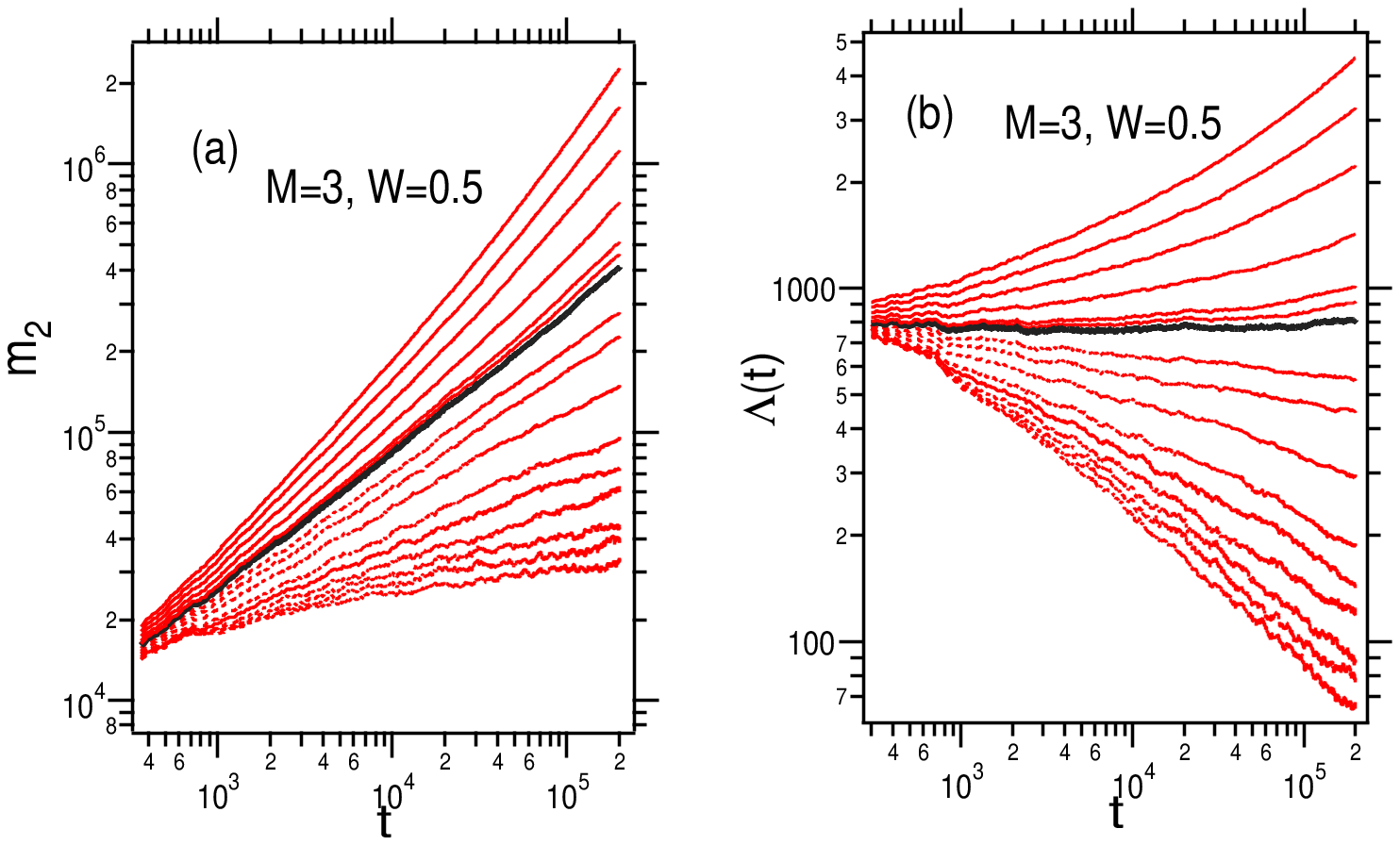}
\hspace{2mm}
\includegraphics[width=8cm]{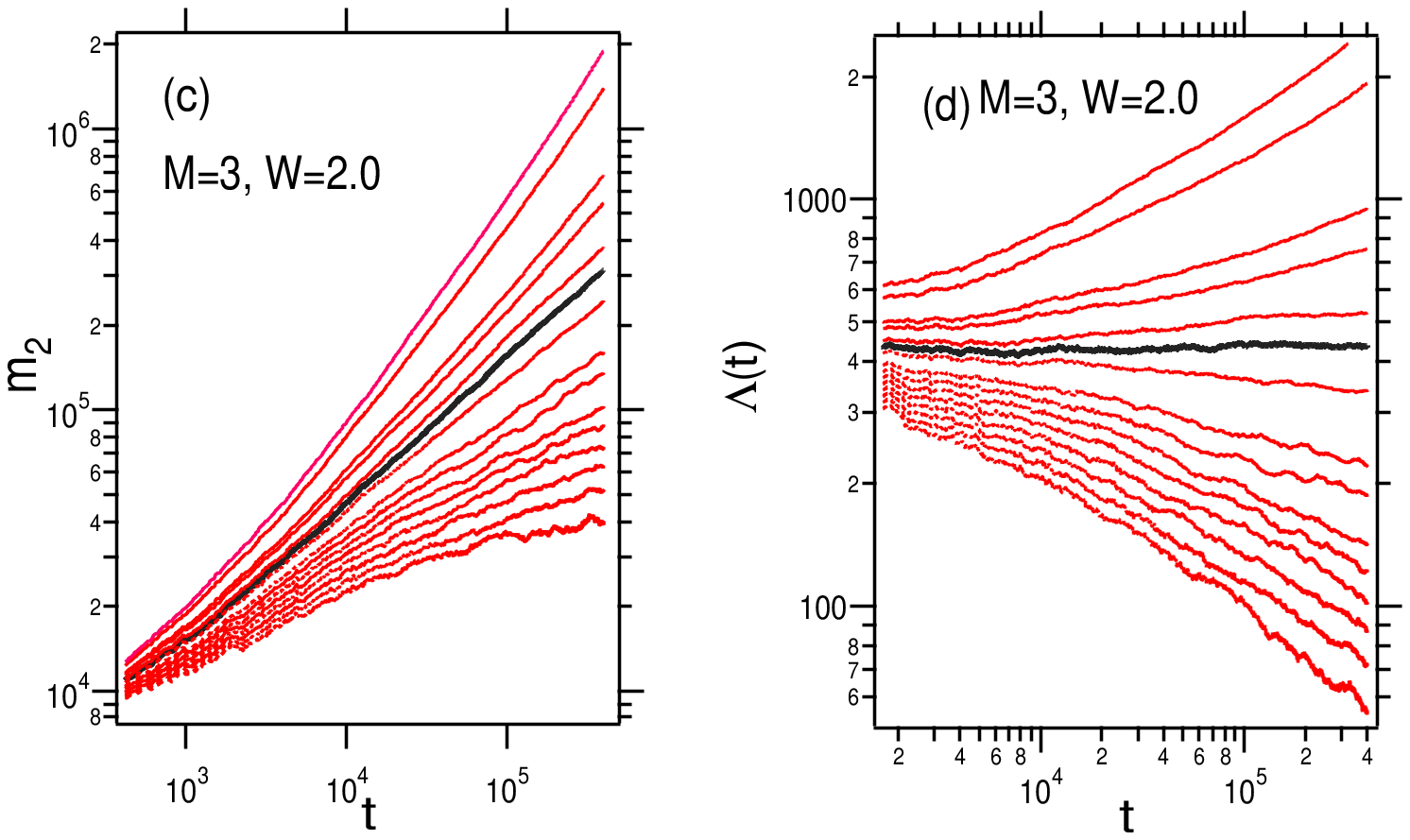}
\hspace{2mm}
\includegraphics[width=8cm]{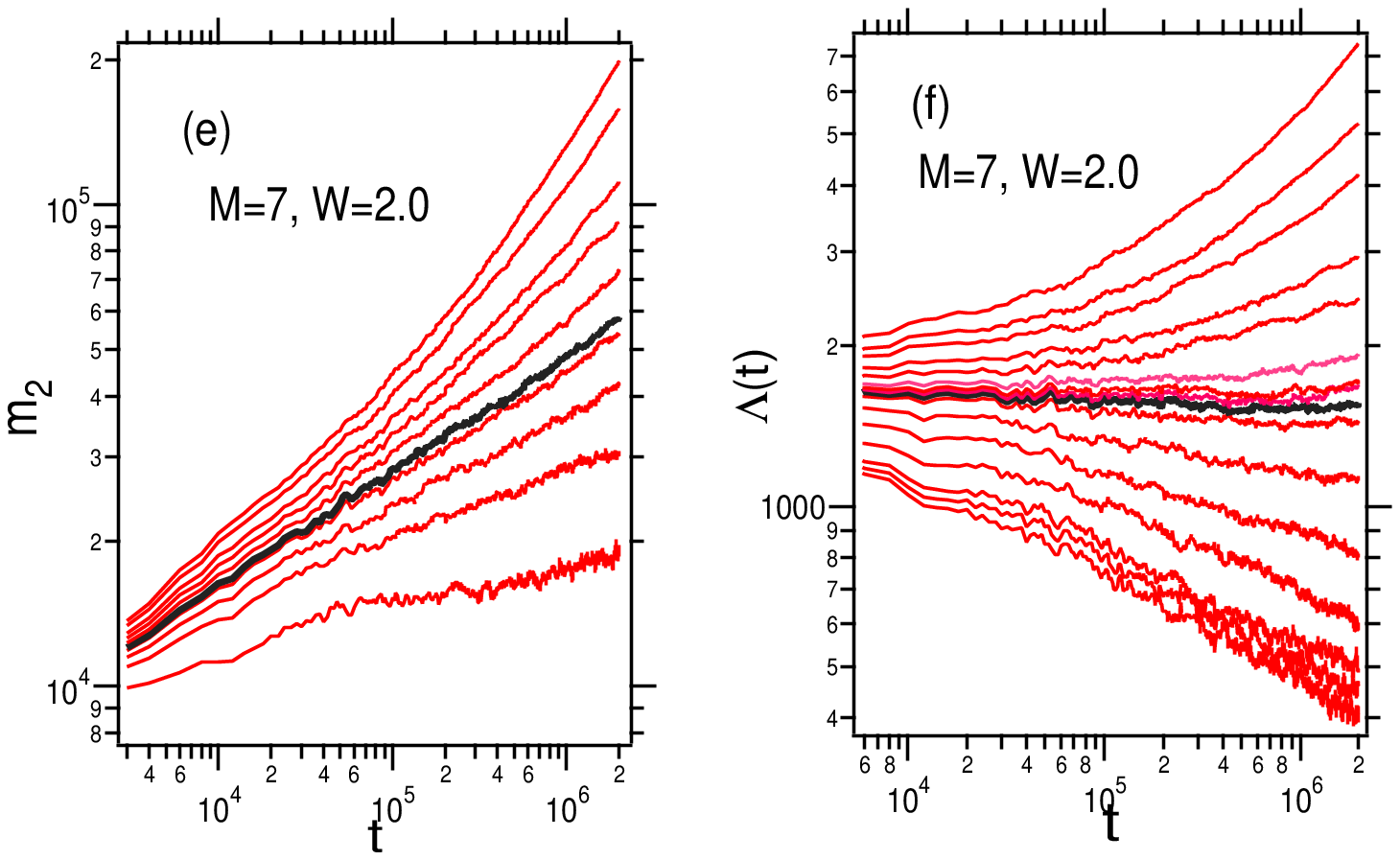}
\hspace{2mm}
\caption{(Color online)
The double-logarithmic plots of (a)$m_2(t)$ 
and (b)the scaled $\Lambda(\eps,t)$ 
as a function of time
for different values of the perturbation strength $\eps$, 
where the diffusion exponent $\alpha$ is determined by 
the least-square-fit for the $m_2(t)$ with the critical case, 
in the trichromatically perturbed AM of $M=3$ with $W=0.5$ 
(c)The same $m_2(t)$ and (d)the scaled $\Lambda(\eps,t)$ 
in the trichromatically perturbed AM of $M=3$ with $W=2.0$.
(e)The same $m_2(t)$ and (f)the scaled $\Lambda(\eps,t)$ 
in the trichromatically perturbed AM of $M=7$ with $W=2.0$.
In the case $M=3$ with $W=0.5$, $\eps_c^{AM}\simeq 0.038$, $\alpha \simeq0.5$.
In the case $M=3$ with $W=2.0$, $\eps_c^{AM}\simeq 0.011$, $\alpha \simeq0.5$.
In the case $M=7$ with $W=2.0$, $\eps_c^{AM}\simeq 0.0068$, $\alpha \simeq0.25$.
We  take $\hbar=0.125$ as the Planck constant for the perturbed AM.
The data near the critical value $\eps_c$ are shown by bold black lines.
}
\label{fig:AM-MSD}
\end{center}
\end{figure}



Next we show the corresponding observations for AM. 
Fig.\ref{fig:AM-MSD} shows the dynamic behavior of AM near the LDT.
According to Eq.(\ref{eq:m2-loc-D-AM}), the disorder strength $W$ of AM 
has the characteristic value $W^*$ beyond which localization characteristics 
change. At fixed $M=3$, the time-dependence of $m_2(t)$ and $\Lambda(t)$
for $W=0.5(<W^*)$ and $W=2.0(>W^*)$ 
are shown in Fig. \ref{fig:AM-MSD}(a),(b) and (c),(d), respectively,  
for various values of $\eps$ increased across the critical value $\eps=\eps_c$ of LDT.
It follows that 
the LDT occurs regardless of the value of $W$. 
The result for $M=7$ is also displayed in Fig.\ref{fig:AM-MSD}(e)(f).
As with the SM, we can see the existence of the LDT and the critical subdiffusion 
with increasing $\eps$.
The critical subdiffusion index $\alpha$ of AM also obeys the
``universal rule''  Eq.(\ref{eq:alpha}) and moreover the
critical value $\eps_c$ depends on $M$ in the same way as the SM.
However the dependence
of $\eps_c$ on the randomness parameter $W$ changes at $W=W^*$.
These properties will be discussed later in detail. 

In the following, the characteristics of LDT are
studied changing the values of control parameters in a wide range.
In SM, we study the change in critical behavior for parameter $K$ 
that controls classical chaos, and Planck constant $\hbar$ 
that controls quantum property, 
whereas AM uses parameter $W$ that controls randomness.
In AM, the size of $\hbar$ is kept at $O(1)$. 
The parameters $K$, $\hbar$ and $W$ are important because they 
decides the localization length $\ell_0$ 
by Eqs.(\ref{eq:m2-loc-D-SM}) and (\ref{eq:m2-loc-D-AM}).

However, in the present study the dependency of LDT 
on the number of the harmonic degrees of freedom $M$ is of particular interest.
Indeed, the change of $M$ is reflected significantly in the characteristics 
of critical subdiffusion index by Eq.(\ref{eq:alpha}), which should also be
reflected in $\eps_c$. 


We are also interested in critical exponents characterizing the divergence of
localization length close to the critical point, but it have been fully discussed 
in the previously published paper [II]. 
Some extensive arguments for this topic is presented in the appendix 
\ref{app:critical-exponent}.

\def\i{i}
\def\j{j}
\def\DZnu{\ovl{\DZ_\i}}

\section{Critical coupling strength of LDT
 in the polychromatically perturbed quantum maps}
\label{sec:epsc}
We focus our attention to the critical value $\eps_c$ of LDT,
which is investigated numerically and compared with theoretical
prediction based upon the SCT.
This is the main part of the present paper.

\subsection{A prediction based on self-consistent theory} 
The critical perturbation strength $\eps_c$ is a quite important parameter featuring
the LDT. The one parameter scaling theory, which is very powerful
for the prediction of critical exponents, is not
applicable to evaluate the critical point.
We use here the SCT for predicting the characteristics of $\eps_c$.

Let $j=0$ assign to the main degrees of freedom of SM and AM, and $j=1,..,M$ to the
$M$ harmonic modes. We regard our systems as $(M+1)-$degrees of freedom one according
to Eqs.(\ref{eq:Maryland_SM}) and (\ref{eq:Maryland_AM}), 
which can be identified with a $d(=M+1)-$dimensional lattice with random and$/$or quasi-periodic on-site potential as is shown in appendix \ref{app:Maryland}. Then we can apply the 
scheme of SCT for the $d-$dimensional disordered lattice system to our system.
Let the frequency-dependent diffusion constant of the $j$-mode be $D_{j}(\omega)$.
The ratio of $D_{j}(\omega)$ to the bare diffusion constant $\DZ_{j}$ is reduced
from 1 by the correction due to the coherent backward scattering, satisfying the relation 
\begin{widetext}
\begin{eqnarray}
\frac{D_{j}(\omega)}{\DZ_{j}} = 1-
C \frac{D_{j}(\omega)}{\DZ_{j}}
\int^{q_{0}^{c}}...\int^{q_{d-1}^{c}}\prod_{k=0}^{d-1}dq_{k}\frac{1}{-i\omega+\sum_{k=0}^{d-1}
D_k(\omega)q_k^2}, 
\label{eq:sce01}
\end{eqnarray}
\end{widetext}
where $C$ is a constant value independent of the parameters.
Note that the integral over $q_k$ has a cutoff $q_{k}^{c}$, which plays a crucial role
\cite{yamada18}.
If we set
\begin{eqnarray}
\label{eq:sce02}
\frac{D_{j}(\omega)}{-i\omega} = \xi_j(\omega)^2,
\end{eqnarray}
then $\lim_{\omega \to 0}\xi_j(\omega)=\ell_j$ becomes the localization length.
In the limit of $\eps=0$, the propagation along the mode $j$ terminates at the localization length 
$\ell_j$. We suppose that the inverse of $\ell_j$ decide the cut-off wavenumber $q_{k}^{c}$, i.e., 
\beq
\label{sce-qcut}
  q_{j}^{c}\sim \ell_{j}^{-1}, 
\eeq
which correctly predicts numerical results of the localization process in the case of $M\leq 2$
\cite{yamada18}. 
As the localization length of the main mode $j=0$ we take $\ell_0$ of 
Eqs.(\ref{eq:m2-loc-D-SM}) and (\ref{eq:m2-loc-D-AM}),
then Eq.(\ref{eq:m2-loc-D}) holds and $\ell_0=\DZ_0$. The diffusion along the harmonic mode
$j$ occurs according to Eq.(\ref{eq:propcolor}), being driven by the force $\hatG(t)$.
Similarly to Eq.(\ref{eq:m2}), the MSD of the harmonic mode $j$ grows as
\beq
\label{eq:sce-color1}
\left<(\haty(t)-\haty(0))^2\right> = \sum_{s\leq t}\DZ_j(s:t), 
\eeq
where
\beq
&& \nn \DZ_j(s:t)= \\
&&~C_j^2\frac{\eps^2}{2M}\sum_{s'=s}^{t}\<\hatG(s')\hatG(s)\>\cos(\omega_j(s'-s))+c.c , 
\eeq
where the average over the initial phase $\phi_{j0}$ is done.

In the case of SM, the force driving the diffusion of the main mode 
$\hatF(t)\propto \sin\hatq$ (Eq.(\ref{eq:m2})) has the same correlation property as that of 
the harmonic mode $\hatG(t) \propto \cos\hatq$. For AM, we also use the same assumption
that the driving force for the harmonic mode ($\hatG(t)=\sum_nv_n|n\>\<n|/\hbar~~(|v_n| \sim O(1))$) 
and that for the main mode ($\hatF(t)=2\sin(\hatp/\hbar)/\hbar=\sum_n(|n\>\<n+1|-|n+1\>\<n|)/(i\hbar)$)
has the same correlation property. Then following the idea of deriving Eq.(\ref{eq:m2-loc}), the 
diffusion of the harmonic mode terminates at the localization time $t_0=\ell_0$ of the main mode
and so the localization length of the mode $j$ is 
\beq
\label{eq:sce-color2}
  \ell_j^2 = \DZ_j\ell_0   
\eeq
by using the initial stage diffusion constant $\DZ_j:=\DZ_j(s=0,t=\infty)$ of the $j$-mode.
Let us define $\kappa_j(\omega):=\frac{\xi_j(\omega)}{\ell_j}$ which is the ratio 
of the enhanced localization length to the localization length.
Then in the self-consistent equation  (\ref{eq:sce01}) the only $j$ dependent
parameter is $D_j(\omega)/\DZ_j$, which is rewritten by using
Eqs.(\ref{eq:sce01}) and (\ref{eq:sce-color2}) as
\beq
\nn \frac{D_j(\omega)}{\DZ_j}=-i\omega \kappa_j(\omega)^2\ell_0.
\eeq
In order that all the equations for $j=0,1,...,d-2, d-1(=M)$ in Eq.(\ref{eq:sce01})
are consistent, $\kappa_j(\omega)$ should be equal and independent of $j$.
By rescaling $q_k'=q_k\xi_k(\omg)$, the integral of Eq.(\ref{eq:sce01}) 
can be approximated as 
the $d$-dimensional spherical integral over the radius $\kappa_k=\kappa_0$. If $\kappa(\omega)$
is much greater than unity assuming that $\eps$ is close to the critical point,
Eq.(\ref{eq:sce01}) is integrated as
\beq
\label{eq:sct-2}
 \frac{D_{j}(\omega)}{\DZ_{j}} 
&=& 1 - \frac{CS_{d}}{(d-2)\prod_{k=1}^{d-1}\ell_k}.
\eeq
$S_d$ denotes the surface area of the $(d+1)$-dimensional sphere of radius unity:
\beq
 S_d =\frac{2\pi^{d/2}}{\Gamma(\frac{d}{2})}.
\eeq

According to Eq.(\ref{eq:sce-color1}) the diffusion constant $\DZ_j$
of the $j(\neq 0)$-mode is the product of the factor $\frac{\eps^2}{2M}C_j^2$
and the time-integral of the correlation function of $\hatG$, which is
the same as that of the driving force $\hatF$ of the main mode, as discussed
above. Therefore, the diffusion constant of  the $j(\neq 0)$mode is related to that
of the main mode as
%
\beq
\label{eq:sce-D0nu}
   \DZ_j = \frac{\eps^2}{2M}C_j^2\DZ_0.
\eeq
Note that $\DZ_0$ is the diffusion constant of isolated main mode
independent of $\eps$ and $M$.

The critical coupling strength $\eps_c$ which makes the l.h.s. of 
Eq.(\ref{eq:sct-2}) zero is given as the condition for the
harmonic mode $j\neq 0$ as follows:
\beq
\label{eq:sce-epsc-PRE}
   \ell_j = \left[\frac{CS_{M+1}}{(M-1)}\right]^{1/M}.
\eeq
From Eqs.(\ref{eq:sce-color2}) and (\ref{eq:sce-D0nu}) $\ell_j$ is proportional to 
$\eps\ell_0$, and the critical coupling strength is
\beq
\label{eq:sce-epsc}
   \eps_c = \frac{c_M}{\ell_0C_j}, 
\eeq
where the parameter $M$ is contained in $c_M=[CS_{M+1}/(M-1)]^{1/M}\sqrt{2M}$. 
If $M\gg 1$ the factor $1/\sqrt{M}$ in $c_M$ cancels with $M^{1/2}$ coming
from the $(M+1)$-dimensional spherical surface area $S_M$, and $\eps_c$ does no longer 
depends upon $M$. This prediction will be compared with the numerical results.

In the case of SM the critical coupling strength is given from 
Eq.(\ref{eq:m2-loc-D-SM}):
\beq
\label{eq:ecSM}
 \eps_c^{SM} \sim 1/\ell_0 \sim \hbar^2/D_{cls}\sim \left( \frac{K}{\hbar} \right)^{-2}~~(K\gg1),
\eeq
whereas, in the case of AM, following Eq.(\ref{eq:m2-loc-D-AM}), 
the critical value changes its dependency upon $W$ at $W=W^*$: 
\beq
\label{eq:ecAM}
   \eps_c^{AM} \sim 1/(\ell_0C_j) \sim  
      \begin{cases}
         W~~~~~~(W<W^*), \\
         \dfrac{W^{*2}}{W}~~~(W>W^*).  
      \end{cases}
\eeq
All the above results are the predictions of the SCT.

\subsection{Numerical characteristics of the critical value for fixed color number}
We summarize in this section the results obtained by numerical simulations
and compare them with the predictions of the SCT.
The dependency of $\eps_c$ on the control parameters except for $M$
is discussed in this section.  

\subsubsection{ The SM }
\begin{figure}[htbp]
\begin{center}
\includegraphics[width=8.8cm]{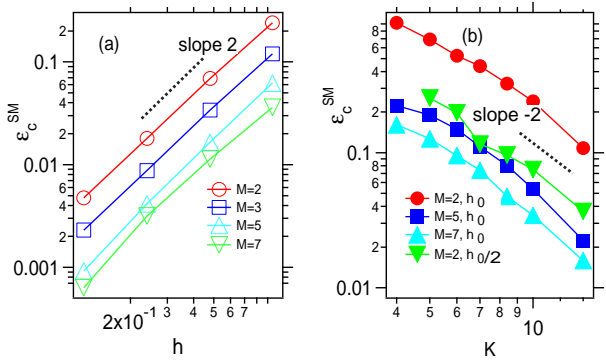}
\caption{(Color online)
(a)The critical perturbation strength $\eps_c^{SM}$ as a function of 
$\hbar$ for the polychromatically perturbed SM ($M=2,3,5$)
with $K=3.1$.
(b)The critical perturbation strength $\eps_c^{SM}$ as a function of 
$K$ for the polychromatically perturbed SM ($M=2,5,7$)
with $\hbar=2\pi 311/2^{13}$, and $M=2$, $\hbar=2\pi 311/2^{14}$.
$\eps_c^{SM} \propto \hbar^{-2}$ and $\eps_c^{SM} \propto K^{2}$
are shown by black broken lines in the panel (a) and (b), respectively.
Note that the axes are in the logarithmic scale.
}
\label{fig:SM-eps-K}
\end{center}
\end{figure}

We first show the critical coupling strength $\eps_c^{SM}$ for SM.
Fig.\ref{fig:SM-eps-K}(a) depicts $\hbar$-dependence of $\eps_c^{SM}$.
Irrespective of the color number $M$ and $K$, the critical strength 
follows evidently the rule $\eps_c^{SM} \propto \hbar^2$.

\beq
\label{eq:epscNUM}
\eps_c^{SM} \sim \left( \frac{K}{\hbar} \right)^{-2}
\eeq
On the other hand, Fig.\ref{fig:SM-eps-K}(b) shows the
dependence upon $K$ with $M$ and $\hbar$ being fixed. It is
strongly suggested that for $K\gg 1$ the critical coupling 
strength obeys the rule $\eps_c^{SM} \propto K^{-2}$ for
the fixed parameters $M$ and $\hbar$ whose values are 
changed over a wide range. Thus we may conclude that the result
of the SCT (\ref{eq:ecSM}) can describe the characteristics
of the critical coupling strength as long as 
two parameters $K$ and $\hbar$ are concerned.

In the case of $M \leq 1$ where the system is localized and there is no LTD,
the characteristics of localization is decided by $K^2/\hbar^2$, which just
means the localization length.  It is quite reasonable that the threshold 
of LDT is decided as $1/\ell_0$. 

In the SCT we suppose a cut-off wavenumber $q_c\sim 1/\ell_j$. 
An another hypothesis is to take the inverse of 
the mean free path $q_c\sim \hbar/K$ \cite{delande}. 
This choice, however, result in the prediction $\eps_c^{SM} \propto K/\hbar$ 
for $M\gg 1$, which contradicts with the numerical results.

\subsubsection{ The AM }
\begin{figure}[htbp]
\begin{center}
\includegraphics[width=7.5cm]{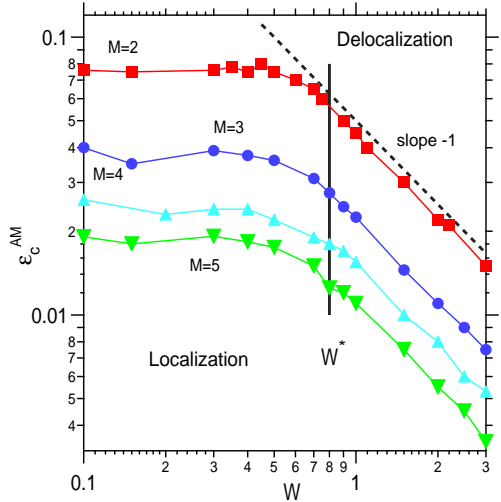}
\caption{(Color online)
The critical perturbation strength $\eps_c^{AM}$ as a function of 
$W$ for the polychromatically perturbed AM  ($M=2,3,4,5$).
$\eps_c^{AM} \propto W^{-1}$ and $W=W^*$
are shown in by dotted black and thick black lines, respectively.
Note that the axes are in the logarithmic scale.
}
\label{fig:c2-phase-1}
\end{center}
\end{figure}

In the case of AM, the critical value $\eps_c^{AM}$
depends upon $W$ as shown by Fig.\ref{fig:c2-phase-1}(a)
for various values of $M$.
The dependence of $\eps_c^{AM}$ upon $W$ changes at $W=W^*$, 
which is consistent with the prediction of the SCT given by Eq.(\ref{eq:ecAM}). 
In particular in the regime $W>W^*$ it is evident that the numerical result 
follows the result of SCT 
\beq
\eps_c^{AM} \sim \frac{1}{W}~~~~(W>W^*)
\label{eq:W-dep1}
\eeq

On the contrary, in the opposite regime $W< W^*$ the numerical
results strongly suggest that
\beq
\eps_c^{AM} \simeq const~~~(W<W^*)
\label{eq:W-dep2}
\eeq
which do not agree with the prediction of the SCT.
Such a tendency persists as $W$ decreases further, and it seems that
$\eps_c^{AM}$ approaches toward a constant depending upon $M$ 
as $W\to 0$.


\begin{figure}[htbp]
\begin{center}
\includegraphics[width=4.05cm]{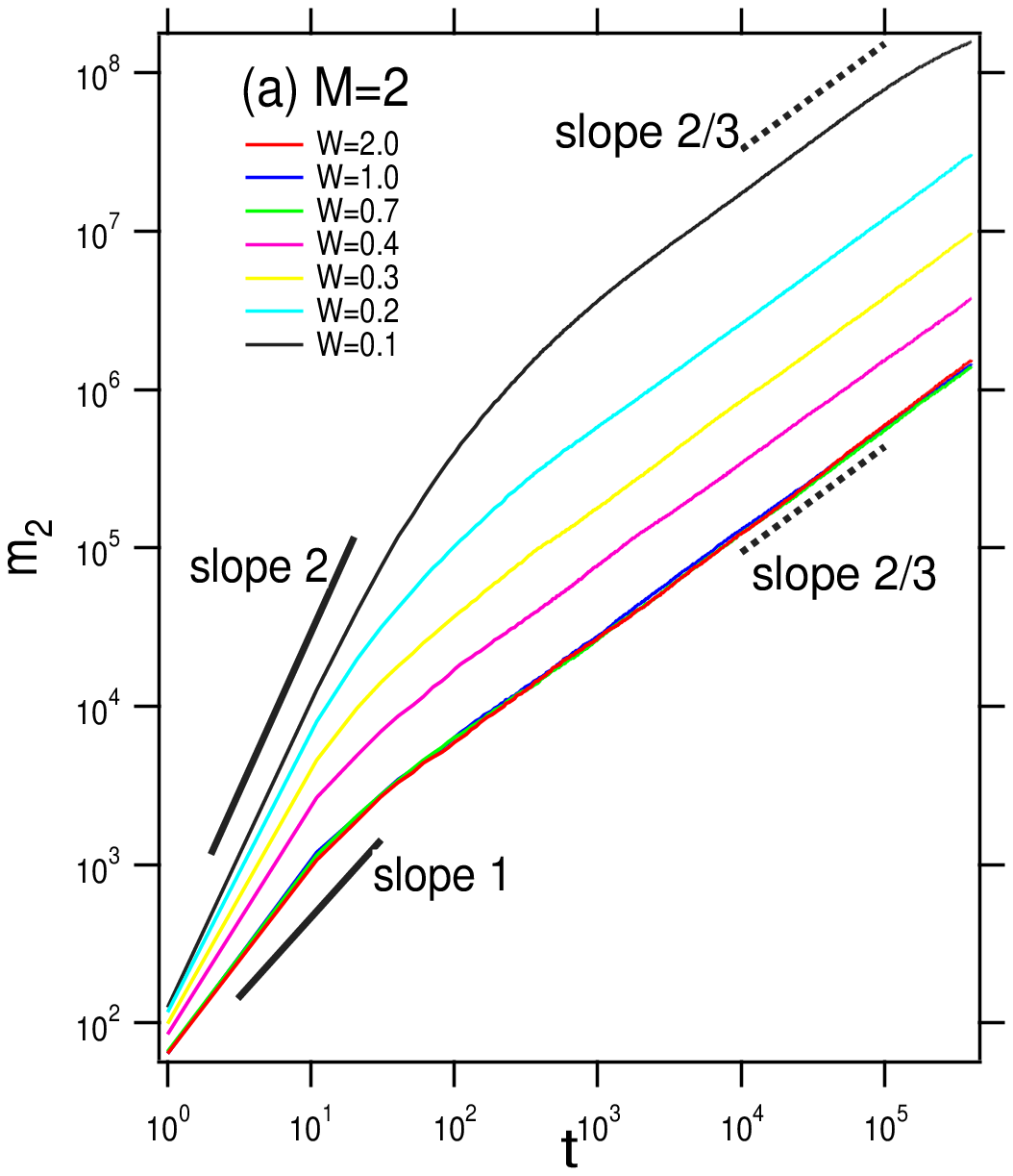}
\hspace{2mm}
\includegraphics[width=4.05cm]{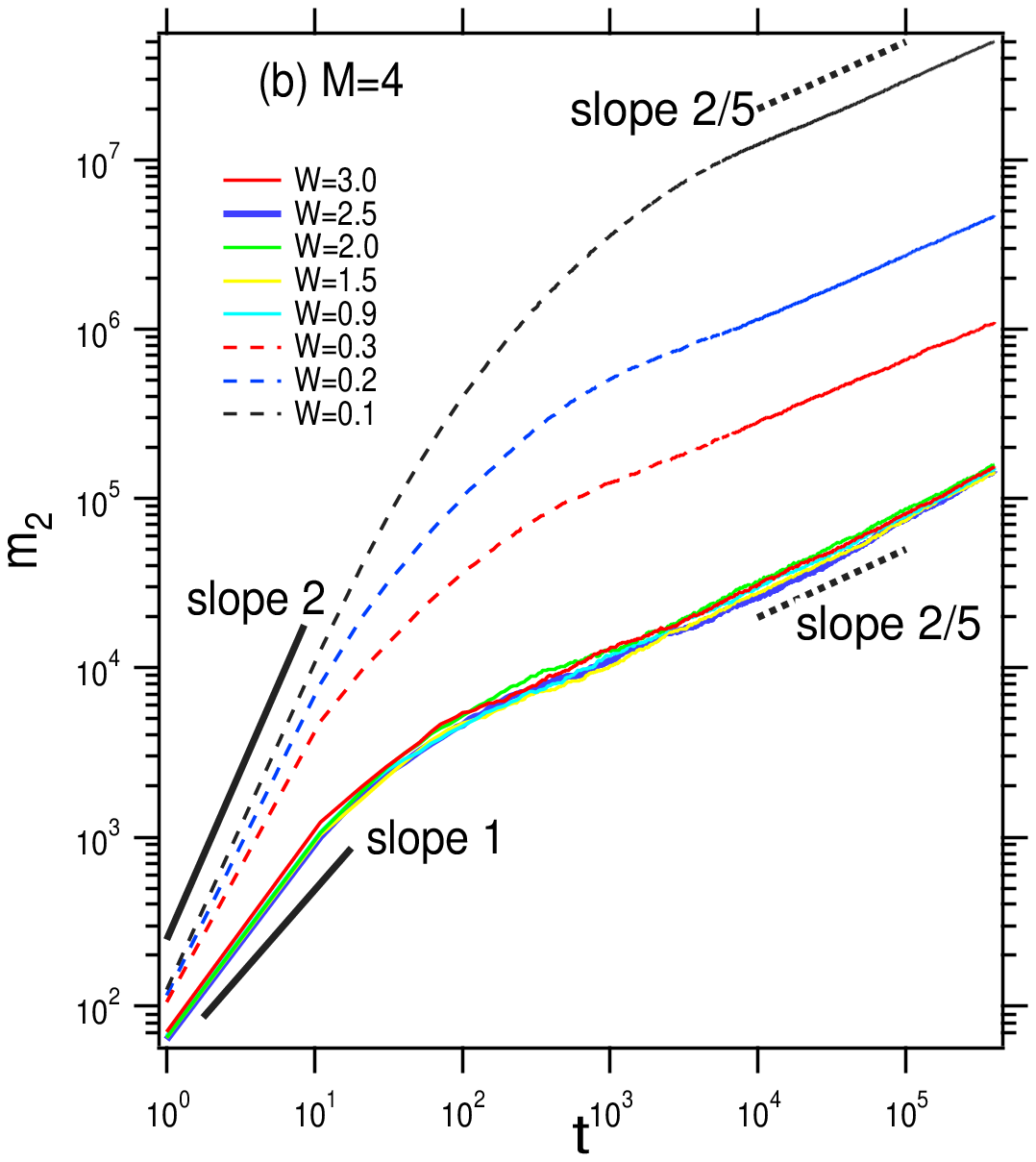}
\caption{(Color online)
The double-logarithmic plots of $m_2(t)$
exhibiting the critical subdiffusion at $\eps_c^{AM}$
are shown for (a)$M=2$ and (b)$M=4$.
Various values of $W$ in the regimes $W<W^*$ and
$W>W^*$ are examined. 
In the regime $W>W^*$ all the curves overlap.
But, as $W\to 0$, $\eps_c^{AM}$ takes the same value
and the ballistic transient motion of slope $2$ become more evident. 
One can see that in the opposite
case $W\gg W^*$ the normal diffusion of slope $1$ first emerges 
before the subdiffusion sets in. 
}
\label{fig:AM-W-epsc}
\end{center}
\end{figure}

\subsubsection{More about AM: the ballistic transient}
The reason why the prediction of SCT fails for 
the AM is tightly connected with a peculiar characteristic of the 
dynamics in the weak $W$ limit of AM.
The basic hypothesis used for deriving Eq.(\ref{eq:W-dep2})
is the the motion of the main mode transiently exhibits
the fully normal diffusion and the harmonic mode follow 
the same transient diffusion process. 
This hypothesis is not, however, correct in 
the weak limit of $W$, because a ballistic motion dominate 
the transient behavior until the scattering occurs at the mean free path length 
$\ell_0$ and makes the motion stochastic. Indeed, in Fig.\ref{fig:AM-W-epsc} 
we can show explicitly how the critical subdiffusion emerges after 
the ballistic transient behavior.

Let us consider the motion of the harmonic mode when the main mode
visits lattice site in a ballistic way until the scattering
at the mean free path $\ell_0$ happens. 
%
%
%
The position of the harmonic mode occurs as
\beq
\haty_j(t) &=&\sum_{s<t}G_j(s)   \nn \\
&=&\sum_{s<t}
\left( \sum_n \frac{\eps W}{\sqrt{M}} v_n|n\>\<n|\right) \sin\omega_j s. 
\eeq
This equation tells that the particle moving at the velocity $V_B$ 
among the lattice sites $|n\>$
causes a randomly switching source proportional to $Wv_n$, which leads to diffusion
of the $j$ oscillator. Then the diffusive motion is expressed by the MSD
\beq
  \<(\haty_j(t)-\haty_j(0))^2\> = \frac{\eps^2 W^2}{M}V_B t, 
\eeq
and so the diffusion constant $\DZ_j= \frac{\eps^2 W^2}{M}V_B$. This motion,
however, terminates the main-mode reaches $\ell_0$ at the time $t_0=\ell_0/V_B$,
and $\ell_j$ should be
\beq
\label{modDZj}
 \ell_j = \sqrt{\DZ_jt_0} = \frac{\eps\sqrt{W^2\ell_0}}{\sqrt{M}}.
\eeq
It is independent of $W$, since $\ell_0\sim W^{-2}$.
Substituting Eq.(\ref{modDZj}) into Eq.(\ref{eq:sce-epsc-PRE}), 
the critical perturbation strength does no longer depends upon $W$, 
which is consistent with the numerical computation.
In the case of SM, we considered an ideal regime such that the diffusion process 
in the classical limit is observed without the coherent dynamical process corresponding 
to the ballistic motion of AM.  However, even in the case of SM, if the coherent motion 
is significant in the classical chaotic diffusion, we need a modification presented above.


\subsection{$M-$dependence of the critical value $\eps_c$}
In the previous subsection the SCT works well for
predicting the characteristics of the critical coupling
strength $\eps_c$ except for the $M$-dependence. However, as is seen
in Figs.\ref{fig:SM-eps-K} and \ref{fig:c2-phase-1}, $\eps_c$
definitely decreases with increase in $M$, and contradict with
the prediction of the SCT. 

\begin{figure}[htbp]
\begin{center}
\includegraphics[width=4.2cm]{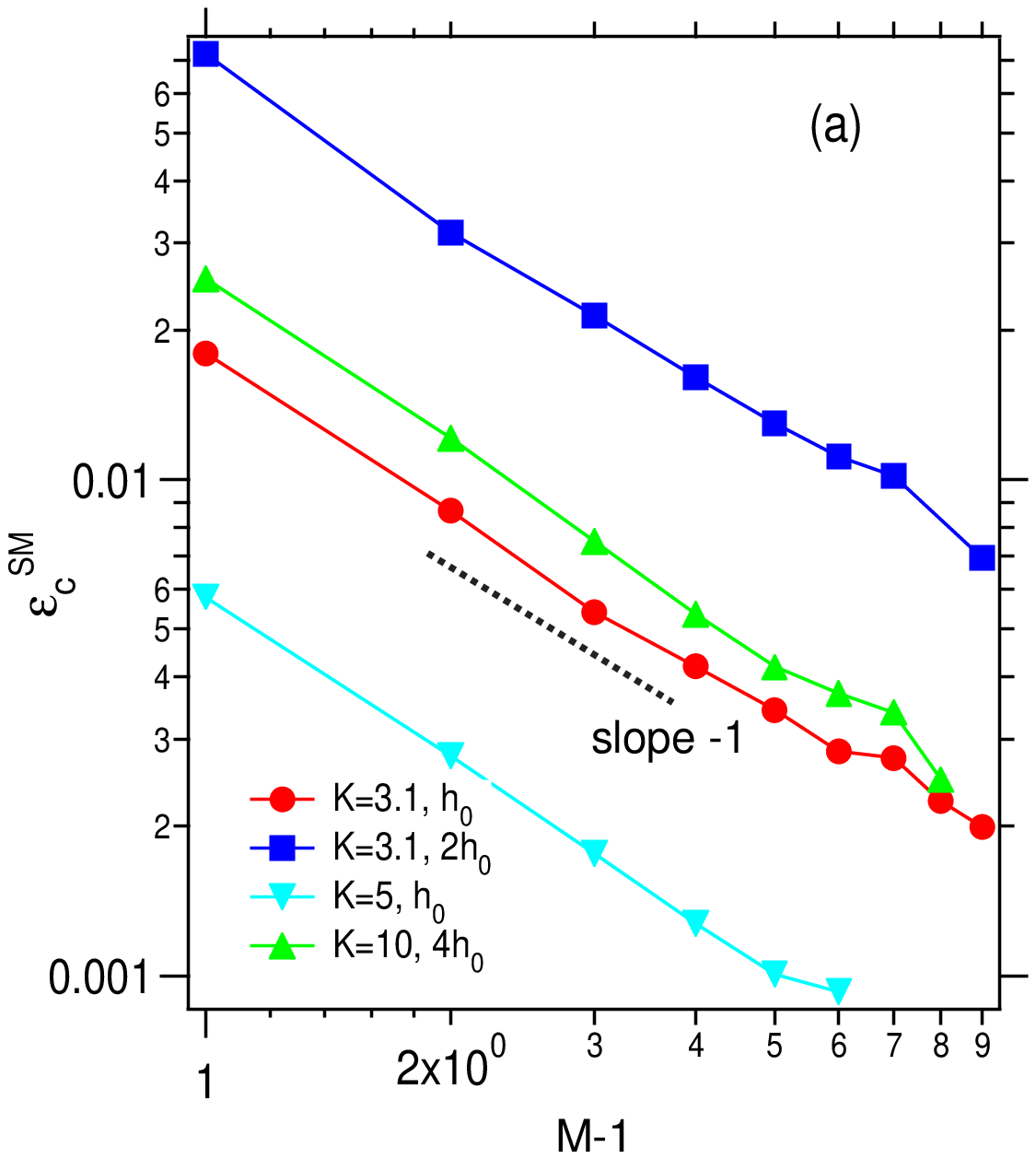}
\hspace{1mm}
\includegraphics[width=4.1cm]{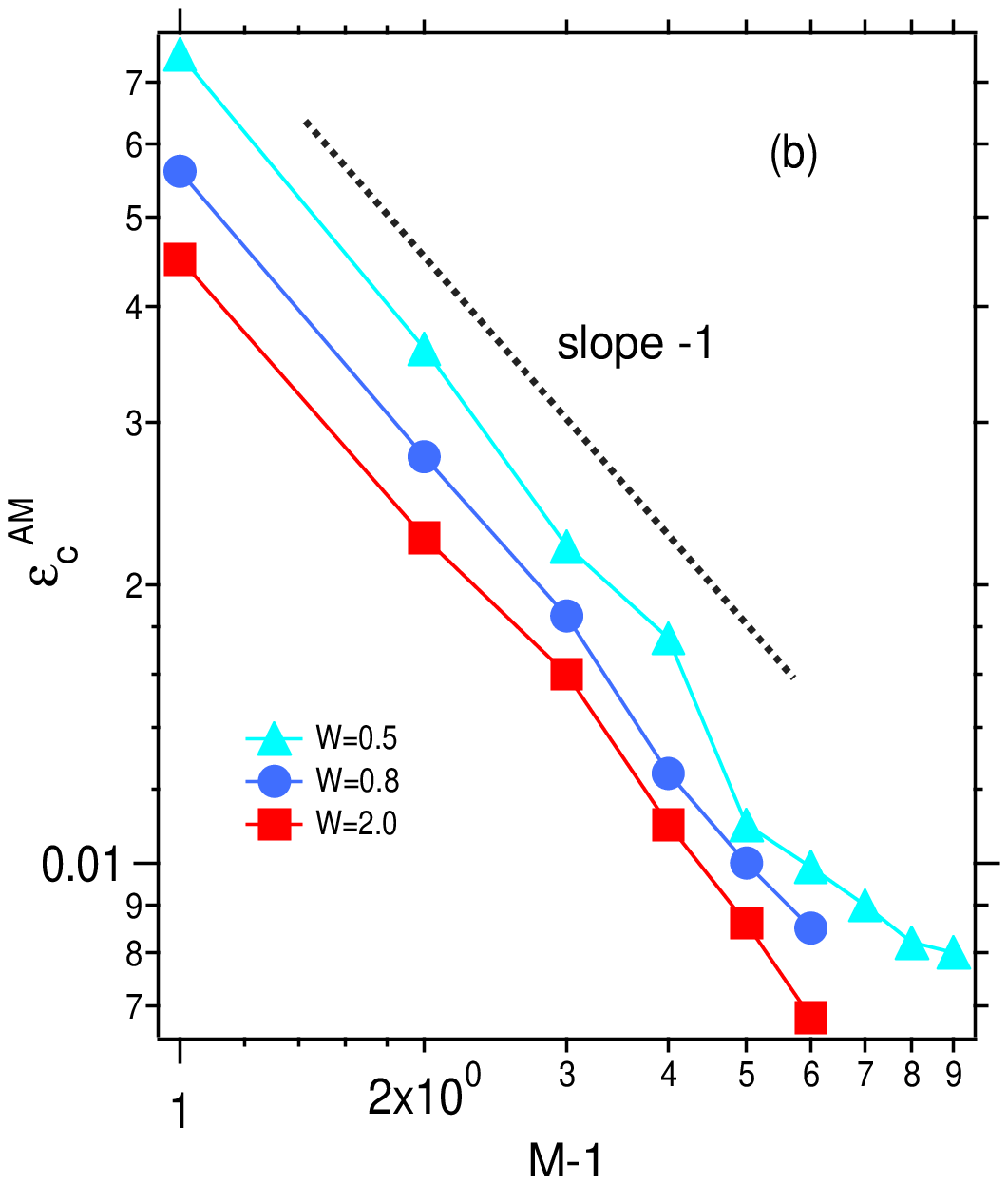}
\caption{(Color online)
(a)The critical perturbation strength $\eps_c^{SM}$ as a function of 
$(M-1)$ for the perturbed SM with $K=3.1$ and $K=5$, 
$\hbar=\hbar_0$.
(b)The critical perturbation strength $\eps_c^{AM}$ as a function of 
$(M-1)$ for the perturbed AM with $W=0.5,0.8,2.0$.
Note that the axes are in the logarithmic scale.
The line with slope $-1$ is shown as a reference.
}
\label{fig:epsc-Mdep}
\end{center}
\end{figure}

With other control parameters such as $K$, $\hbar$ for SM and $W$ for AM being fixed, 
all the numerical results are well fitted by the empirical rule for both SM and AM:
\beq
\eps_c \propto \frac{1}{(M-1)}
\label{eq:eps_cvsM}   
\eeq
as is demonstrated in Fig.\ref{fig:epsc-Mdep}.
Note that divergence at $M=1$ agrees with the absence of LDT in monochromatically 
perturbed SM and AM. (The log-log plot of $\eps_c$ vs $M$ does not form
fine straight curves like those displayed in Fig.\ref{fig:epsc-Mdep}.)
The approach of $\eps$ to zero for $M\to \infty$
means that the localization is destroyed to turn into a normal diffusion by 
the noise with an arbitrary small amplitude.

%

\begin{figure}[htbp]
\begin{center}
\includegraphics[width=7.0cm]{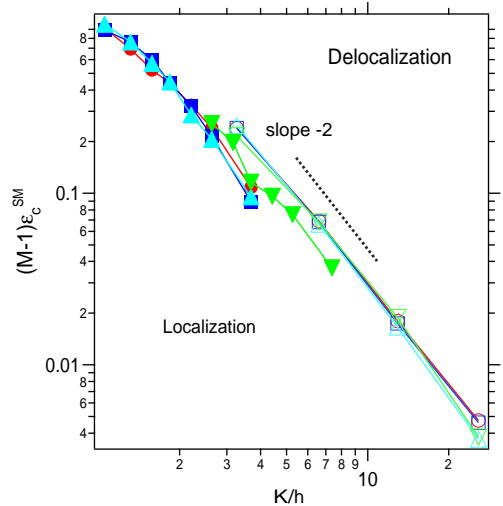}
\caption{(Color online)
The phase diagrams and the critical values $(M-1)\eps_C^{SM}$ in the plane 
$((M-1)\eps,~K/\hbar)$ for the SM. Data with various values 
of $M$, $K$ and $\hbar$ are plotted.
The data plotted by empty circles, squares and crosses, 
which are on a common line in the lower side, 
are the data in Fig.\ref{fig:SM-eps-K}(a). They have the same $K=3.1$,
which is not $K \gg 1$, and so the common line slightly shifts 
from the curves of other data.
The line with slope $-2$ is shown as a reference.
}
\label{fig:SM-M-K-hbar-2}
\end{center}
\end{figure}

Having the rule of Eq.(\ref{eq:eps_cvsM}) in mind, we reorganize our numerical 
results by plotting $(M-1)\eps_c$ vs other parameters. We replot
the data in Fig.\ref{fig:SM-eps-K}(a) and (b) by assigning the vertical
axis to $\eps_c (M-1)$ and the horizontal axis to $K/\hbar$.
All the data points are on a unified single master curve,
which implies the rule
\beq
\label{eq:eps_c_final_SM}
     \eps^{SM}_c \propto \frac{K^2}{\hbar^2(M-1)}
\eeq
exits. A slight discrepancy exits between upper side data and the lower side data.
Its origin will be the fact that the data of the upper side belongs to smaller $K$ regime 
for which significant deviation from the relation $\ell_0 \propto K^2/\hbar^2$ 
occurs. (See caption of the Fig.\ref{fig:SM-M-K-hbar-2}.)

\begin{figure}[htbp]
\begin{center}
\includegraphics[width=7.5cm]{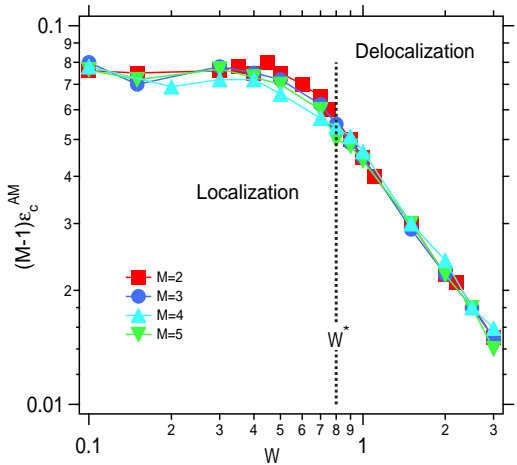}
\caption{(Color online)
The phase diagrams in a plane $((M-1)\eps, W)$ 
for the polychromatically perturbed AM ($M=2,3,4,5$).
$W=W^*$ is shown by dotted black line. 
Note that the axes are in the real scale.
}
\label{fig:c2-phase-1-2}
\end{center}
\end{figure}
%
%
The same plot for the AM on the $((M-1)\eps, W)$ plane is shown
in Fig.\ref{fig:c2-phase-1-2}, which manifests that 
almost all the data are on a single master curve
irrespective of $W<W^*$ or $W>W^*$, which implies the rule
\beq
\label{eq:ec_final_AM}
   \eps_c^{AM} \propto       
         \begin{cases}
         \dfrac{1}{M-1}~~~~~~~(W<W^*), \\
         \dfrac{1}{W(M-1)}~~~(W>W^*).  
      \end{cases}
\eeq
The behavior of the subdiffusion index $\alpha=2/(M+1)$,
which can not be explained by SCT of the localization, seems to be coordinate with
the approach of $\eps_c$ to $0$ with increasing $M$. 
The SCT overestimates $\eps_c$. Indeed,
the second term in the r.h.s. of Eq.(\ref{eq:sce01}),
which evaluates the reduction of the diffusion constant
from the ideal diffusion rate by the backscattering
effect, seems to overestimated. Roughly speaking, this 
integral yields the surface area $S_{M+1} \sim {M}^{-M/2}$, 
which cancels with the normalization factor $1/\sqrt{M}$
and takes off the $M$-dependence from $\eps_c$.
If the surface factor is replaced by a further smaller one
\beq
  S_M'=\frac{1}{M!}S_M,
\eeq
the SCT succeeds in predicting all 
the characteristics of the critical coupling strength.
This replacement means that the $M$-harmonic degrees
of freedom is indistinguishable, but we could not
explain the origin of the above reduction. 


\section{Summary and discussion}
\label{sec:summary}
We investigated the localization-delocalization transition (LDT) of 
the SM (standard map) and the AM (Anderson map) which 
are dynamically perturbed by polychromatically
periodic oscillations for the initially localized quantum wavepacket. 

In the SM and AM, 
for number of colors $M$ more than two, the LDT always takes place with increase
in the perturbation strength $\eps$, and the critical exponents at the critical point
decrease with $M$. In particular, the critical diffusion exponent decreases as 
$\alpha \simeq 2/(M+1)$ in accordance with the prediction of one-parameter 
scaling theory (OPST).

In the present paper, we paid particular attentions to
the dependence of the critical perturbation strength $\eps_c$ 
upon the control parameters. If the number of color $M$ is
fixed, the control parameter dependencies are well predicted by the 
self-consistent theory (SCT) of the localization 
for both SM and AM if basic hypothesis
are properly modified.
On the other hand, the SCT predicts that 
$\eps_c$ does not depends on $M$, while numerical results reveal that
$\eps_c$ reduces drastically as $\eps_c\propto 1/(M-1)$ with an increase of $M$.

The LDT leading to the normal diffusion is a 
decoherence transition, which is basically originated by the entanglement
among wavefunctions spanning the $(M+1)-$dimensional Hilbert space.
Such an entanglement induces the drastic decrease of $\alpha$ and $\eps_c$
with increase of $M$. If $M+1$ can be identified with the spatial
dimension $d$, as is suggested by the Maryland transform, then can we
expect such a steep dependence of critical properties on $d$ for 
Anderson transition in $d$-dimensional disordered
lattice? This is a quite interesting question
\cite{footnote-epsc}.

Owing to such a decrease of threshold $\eps_c$, the polychromic perturbation
is identified with a white noise in the limit $M\to \infty$,
and it can destroy the localization at an arbitrary small amplitude.


It is also a quite interesting problem how such characteristic critical
behaviors are observed for the dynamical delocalization of the time-continuous 
system \cite{yamada99}, 
which shares much common nature with the present AM in the limit 
$W\to 0$. In particular, whether the limiting behavior  $\eps_c\to$ const. in the regime of
$W\ll 1$ is intrinsic and is not due to the discreteness of time evolution 
is an important problem \cite{yamada20}.


\appendix

\section{Malyland transform and tight-binding representation}
\label{app:Maryland}
We consider an eigenvalue equation 
\beq
   \hatU_{aut}|u\> =\e^{-i\gamma}|u\>
\label{eq:eigen-value-problem}
\eeq
for the time-evolution operator of the Hamiltonian (\ref{eq:Haut}),
\beq
  \hatU_{aut}=\e^{-i\hatA}\e^{-i\hatB}\e^{-i\hatC},
\eeq
where $\gamma$ and $|u\>$ are the quasi-eigenvalue and quasi-eigenstate.

For the SM, 
\beq
\begin{cases}
    \e^{-i\hatA} =\e^{-\frac{i}{\hbar}[T(\hatp)+\sum_j^M\omega_j \hatJ_j]},\\
    \e^{-i\hatB} =\e^{-\frac{i}{\hbar}\eps \hatV(q) 
\frac{\eps}{\sqrt{M}}\sum_j^M\cos\phi_j},\\
    \e^{-i\hatC} =\e^{-\frac{i}{\hbar}V(\hatq)}.
\end{cases}
\eeq
The eigenvalue equation we take the representation using
eigenstate $|m\> (m\in{\Bbb Z})$ of momentum $\hatp$ and the action eigenstate 
$\{|m_1\>,...,|m_M\> \} (m_i\in{\Bbb Z})$ of the $M$ number of 
 $J$-oscillators as $u(m,m_1,...,m_M)=(\<m|\otimes \<m_1,...,m_M|)|u\>$.
Then by applying the Maryland transform, the eigenvalue equation can be mapped into 
the following $(M+1)$-dimensional 
tight-binding system with aperiodic and singular on-site potential:
\begin{widetext}
\beq
\label{eq:Maryland_SM}
& & \tan \left[ \frac{\hbar^2m^2/2+\hbar \sum_j^Mm_j\omega_j}{2\hbar}-\frac{\gamma}{2} \right] u(m,m_1,...,m_M)+ \nn \\ 
& & \sum_{m',m_1',...,m_M'}\<m,m_1,...,m_M|\hat{t}_{SM}|m',m_1',...,m_M'\>u(m',m_1',...,m_M')=0,
\eeq
where the transfer matrix element is
\beq
& & \<m,m_1,...,m_M|\hat{t}_{SM}|m',m_1^{'},...,m_M^{'}\>  \nn \\
&=& \frac{1}{(2\pi)^{M+1}}\int_0^{2\pi} ... \int_0^{2\pi}  dq d\phi_1... d\phi_M
\e^{-i(m-m')q}\e^{i\sum_{j}^{M}(m_j-m_j')\phi_j} 
\tan \left[\frac{K\cos q (1+ \frac{\eps}{\sqrt{M}} \sum_{j}^{M}\cos\phi_j)}{2\hbar}  \right].
\eeq
\end{widetext}

On the other hand, for the polychromatically perturbed AM, using 
\beq
\begin{cases}
\e^{-i\hatA} =\e^{-\frac{i}{\hbar} (Wv(\hatq)+\sum_j^M\omega_i \hatJ_j)}, \\
\e^{-i\hatB} =\e^{-\frac{i}{\hbar} v(\hatq)\frac{\eps W}{\sqrt{M}}\sum_j^M\cos\phi_j}, \\
\e^{-i\hatC} =\e^{-\frac{i}{\hbar} 2\cos (\hatp/\hbar)}
\end{cases}
\eeq
 we can also obtain the following 
$(M+1)-$dimensional tight-binding expression:
\begin{widetext}
\beq
\label{eq:Maryland_AM}
& & \tan \left[ \frac{Wv_n+\hbar\sum_j^Mm_j\omega_j}{2\hbar}-\frac{\gamma}{2}
 \right]u(n,m_1,...,m_M)+ \nn \\
& & \sum_{n',m_1',...,m_M'} \<n,m_1,...,m_M|\hat{t}_{AM}|n',m_1',...,m_M'\>u(n',m_1',...,m_M')=0, 
\eeq
where the transfer matrix element is
\beq
& & \<n,m_1,...,m_M|\hat{t}_{AM}|n',m_1^{'},...,m_M^{'}\>  \nn \\
&=&  \Braket{n,m_1,...,m_M|
i\frac{
e^{-i  \frac{\eps W}{\sqrt{M}}v_n (\sum_{i}^{M}\cos\phi_i)/\hbar}-
e^{i2\cos(\hatp/\hbar)/\hbar}
}
{e^{-i\frac{\eps W}{\sqrt{M}}v_n (\sum_{i}^{M}\cos\phi_i) /\hbar}+
e^{i2\cos(\hatp/\hbar)/\hbar}
}
| n',m_1^{'},...,m_M^{'} }.
\eeq
\end{widetext}
The $n$ denotes  one-dimensional disorder site of the AM.
In this representation, the effect of the disorder strength $W$ of the diagonal term 
 saturates at $W^*(=2\pi \hbar)$ and 
 increasing beyond $W^*$ does not affect the diagonal disorder.
Also, it can be seen that the effect of the perturbation is embedded 
in the off-diagonal term representing hopping in the form of $\eps W $
for $W>W^*$.
For this reason, the critical perturbation strength indicates the $W-$dependence
in Eq.(\ref{eq:W-dep1})  when $W>W^*$.

It follows that the $(M+1)-$dimensional tight-binding models
of the SM and AM have singularity of the on-site energy caused by 
tangent function and long-range hopping caused by kick.
However, in the case of $\eps \neq 0$, the evaluation of matrix elements is not easy
since the stochastic quantity $v_n$ is contained in addition to both operators  
$\hatq$ and $\hatp$.

\section{One-parameter scaling theory and diffusion exponent}
\label{app:OPST}
In the long-time limit ($t \to \infty$), we can predict asymptotic behavior of MSD as
\beq
  m_2(t) \sim
\begin{cases}
           \xi^2  & ( \eps <\eps_c)  \\ 
           Dt  & ( \eps>\eps_c),
\end{cases}
\eeq
for the localized ($\eps<\eps_c$) and delocalized regime ($\eps>\eps_c$), respectively.
Here $D$ and $\xi$ denote the diffusion coefficient and localization length, respectively.
In the vicinity of LDT $\eps \simeq \eps_c$, with two critical exponents $\nu$ and $s$, 
we assume 
\beq
\begin{cases}
            D \sim (\eps-\eps_c)^s  & ( \eps>\eps_c) \\ 
           \xi \sim (\eps_c-\eps)^{-\nu}  & ( \eps <\eps_c,).
\label{eq:d-xi}
\end{cases}
\eeq
The exponents satisfy Wegner relation 
\beq
 s=(d-2)\nu
\eeq
where $d$ is spatial dimension \cite{wegner76}.

We can use the following scaling hypothesis
\beq
m_2(t)=a^2F_1(L_t/a, \xi/a), 
\eeq
with two-variable scaling function $F_1(x_1,x_2)$.
Here an unique characteristic length  $L_t$ associated with dynamics as 
\beq
L_t \sim t^\sigma,
\eeq
where $\sigma$ is a dynamical exponent.
If we set $a=\xi$, then  $m_2$ scales like 
\beq
m_2(t) &=& \xi^2F_1(t^\sigma/\xi, 1), \\ 
&=& t^{2\sigma}F_2(t^{\sigma/\nu}(\eps-\eps_c)),
\eeq
where $F_2(x)$ is a one-variable scaling function.
A relation 
\beq
2\sigma+\frac{\sigma s}{\nu}=1,
\eeq
must be satisfied to recover the condition (\ref{eq:d-xi}).
Using Wegner relation it follows
\beq
 \sigma=\frac{1}{d}.
\eeq
Therefore, at the critical point $\eps=\eps_c$ of LDT, the MSD shows 
subdiffusion 
\beq
m_2(t) \sim t^{\alpha}.
\eeq
with the diffusion exponent 
\beq
 \alpha=\frac{2}{d}=\frac{2}{M+1}.
\eeq

\section{Critical localization exponents of LDT in the polychromatically perturbed 
quantum maps}
\label{app:critical-exponent}
In this appendix, the finite-time scaling analysis of the LDT 
by using MSD $m_2(t)$ and the $M-$dependence of the critical exponent
in the perturbed SM and AM are shown.
However, note that pursuing $\nu$ by numerical calculations with high accuracy 
is not the purpose of this paper.

First, let us consider the following quantity 
\beq
\Lambda_s(\eps,t)=\frac{\Lambda(\eps,t)}{\Lambda(\eps_c,t)}-1
\eeq
as a scaling variable.
For $\eps > \eps_c$, 
the  $\Lambda_s$ increases and the wave packet delocalizes with time.
On the contrary, for  $\eps < \eps_c$,  $\Lambda_s$ decreases with time and the
wavepacket turns to the localization.
Around the LDT point of the perturbed cases by $M$ modes,
the localization length $\xi$ is supposed to diverge 
\beq
\xi \sim |\eps_c-\eps|^{-\nu}
\eeq
as $\eps \to \eps_c$ for the localized regime $\eps \leq \eps_c$.
$\nu$ of LDT is the critical localization exponent characterizing divergence  of the 
localization length and depends on the number of modes $M$, 
but after that, the subscript $M$ is abbreviated for simplicity of the notation.

For $\Lambda_s(t)$, it is assumed that in the vicinity of this LDT  
one-parameter scaling theory (OPST) is established
as the parameter is the localization length $\xi(\eps)$.
Then,  $\Lambda_s(t)$ can be expressed as,
\beq
\Lambda_s(\eps,t) &=& F(x),
\label{eq:real-scale-0}
\eeq
where 
\beq
x=(\eps_c-\eps)t^{\alpha/2\nu}.
\eeq
$F(x)$ is a differentiable scaling function and $\alpha$ is the diffusion index.
Therefore, $F(x)$ is expand around the critical point as follows: 
\beq
F(x)=F(0)+C_1(t)(\eps_c-\eps)+C_2(t)(\eps_c-\eps)^2+....
\eeq
, and
\beq
C_1(t)  \propto  t^{\alpha/2\nu}.
\label{eq:critical-exponent}
\eeq
As a result, the critical exponent $\nu$ of LDT can be determined using data obtained by numerical calculation and the above relation.
If we use the  $\nu$ and $\alpha$, 
 we can ride $\Lambda_s$ for various $\eps$ on a smooth function 
by shifting the time axis to $x$.
This is consistent with formation of the scaling hypothesis.

\begin{figure}[htbp]
\begin{center}
\includegraphics[width=7.0cm]{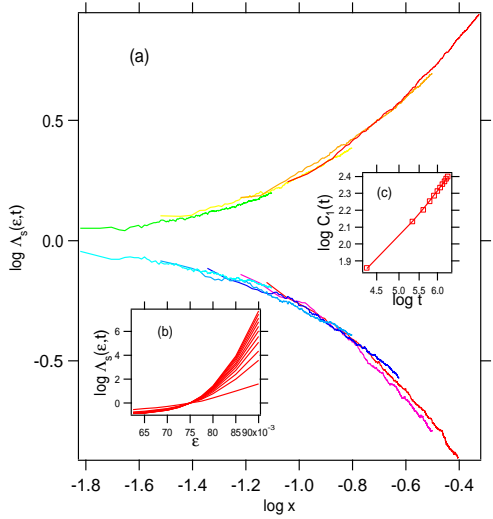}
\caption{(Color online)
The results of the critical scaling analysis for 
trichromatically perturbed SM ($M=3$) with $K=3.1$ and $\hbar=\hbar_0$.
(a)The scaled variable $\Lambda_s(\eps,t)$ 
as a function of $x=|\eps_c^{SM}-\eps|t^{\alpha/2\nu}$
 for some values of $\eps$.
(b)The scaled MSD $\Lambda_s(\eps,t)$ with $\alpha \simeq 0.46$ 
as a function of $\eps$  for some pick up times.
The crossing point is $\eps_c^{SM} \simeq 0.13$.
(c)$C_1(t)$ as a function of $t$.
The critical exponent $\nu \simeq 0.95$ is determined by a scaling relation 
Eq.(\ref{eq:critical-exponent}) by the least-square fit.
}
\label{fig:c3-nu-sm-1}
\end{center}
\end{figure}

\begin{figure}[htbp]
\begin{center}
\includegraphics[width=7.0cm]{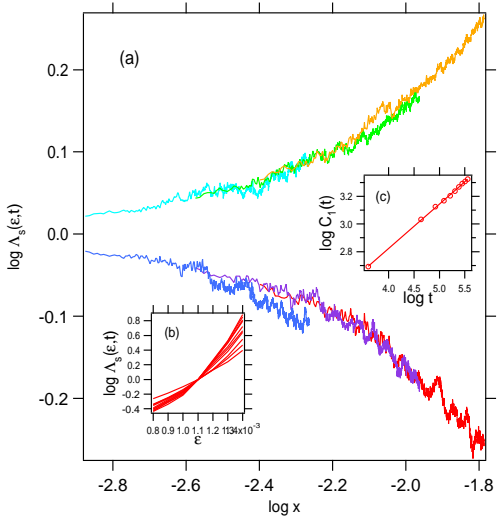}
\caption{(Color online)
The results of the critical scaling analysis for 
polychromatically perturbed SM ($M=7$) with $K=3.1$ and $\hbar=\hbar_0$.
(a)The scaled variable $\Lambda_s(\eps,t)$ 
as a function of $x=|\eps_c^{SM}-\eps|t^{\alpha/2\nu}$
 for some values of $\eps$.
(b)The scaled MSD $\Lambda_s(\eps,t)$ with $\alpha \simeq 0.25$ 
as a function of $\eps$  for some pick up time $t_m$.
The crossing point is $\eps_c^{SM} \simeq 0.018$.
(c)$C_1(t)$ as a function of $t$.
The critical exponent $\nu \simeq 0.35$ is determined by a scaling relation 
Eq.(\ref{eq:critical-exponent}) by the least-square fit.
}
\label{fig:c3-nu-sm-2}
\end{center}
\end{figure}

Figure \ref{fig:c3-nu-sm-1} shows the scaling curve constructed by the time-dependent
data at various $\eps$ near $\eps_c$ in SM of $M=3$
with $K=3.1$, $\hbar=\hbar_0$.
Figure \ref{fig:c3-nu-sm-1}(b) is a plot of $\Lambda_s(\eps,t)$ 
as a function of $\eps$ at several times $t$, 
and this crosses at the critical point $\eps_c$.
Also, Fig. \ref{fig:c3-nu-sm-1}(c) shows $C_1(t)$
as a function of $t$, and the critical exponent $\nu$ is determined 
by best fitting the slope, and the scaling curve $F(x)$ is displayed 
in \ref{fig:c3-nu-sm-1}(a) using the critical values.
It is well scaled and demonstrates the validity of OPST.
Further, Fig.\ref{fig:c3-nu-sm-2} displays 
the result of the finite-time scaling analysis for
polychromatically perturbed SM ($M=7$) with $K=3.1$ and $\hbar=\hbar_0$.
For any number of colors $M$, the LDT is well scaled 
against perturbation strength changes, 
suggesting that LDT can be described fairly well within the OPST framework.

\begin{figure}[htbp]
\begin{center}
\includegraphics[width=7.0cm]{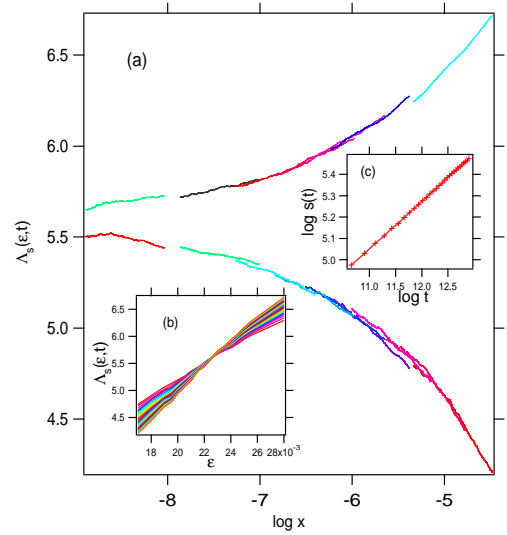}
\caption{(Color online)
The results of the critical scaling analysis for 
dichromatically perturbed AM ($M=2$) with $W=2.0$($>W^*$).
(a)The same scaled MSD $\Lambda_s(\eps,t)$ 
as a function of $x=\xi_0|\eps_c^{AM}-\eps|^{-\nu}t^{\alpha/2\nu}$ for some values 
of $\eps$, where $\xi_0$ is the localization length in the unperturbed case.
(b)The scaled $\Lambda_s(\eps,t)$ with $\alpha=0.65$ 
as a function of $\eps$  for some pick up times.
The crossing point is $\eps_c^{AM} \simeq 0.0225$.
(c)$s(t)$ as a function of $t$.
The critical exponent $\nu \simeq 1.48$ is determined by a scaling relation 
Eq.(\ref{eq:s(t)}) by the least-square fit.
}
\label{fig:c2-nu-w20}
\end{center}
\end{figure}

In Fig.\ref{fig:c2-nu-w20}, we show result of finite-time scaling 
analysis for AM of $M=2$ with $W=2.0(>W^*)$.
The method used here is the same as that used in the paper [2]
for  AM of $M=5$ with $W=0.5(<W^*)$.
We choose the following quantity as a scaling variable
\beq
\Lambda_s(\eps,t)=\log \Lambda(\eps,t).
\eeq

Figure \ref{fig:c2-nu-w20}(b) shows a plot of $\Lambda_s(t)$ 
as a function of $\eps$ at several times $t$, and 
it can be seen that this intersects at the critical point $\eps_c$.
In addition, Fig.\ref{fig:c2-nu-w20}(c) shows a plot of 
\beq
s(t)=\frac{\Lambda_s(\eps,t)-\Lambda_s(\eps_c,t)}{|\eps_c-\eps|}  \propto  t^{\alpha/2\nu} 
\label{eq:s(t)}
\eeq
as a function of $t$, and the critical localization exponent $\nu$ 
is determined by best fitting this slope.
In Fig.\ref{fig:c2-nu-w20}(a), we plot $\Lambda_s$ as a function of $x$ for 
different values of $\eps$ by using the obtained the critical exponent $\nu$.
Similar results to case in the paper [2] is obtained. 

\begin{figure}[htbp]
\begin{center}
\includegraphics[width=7.0cm]{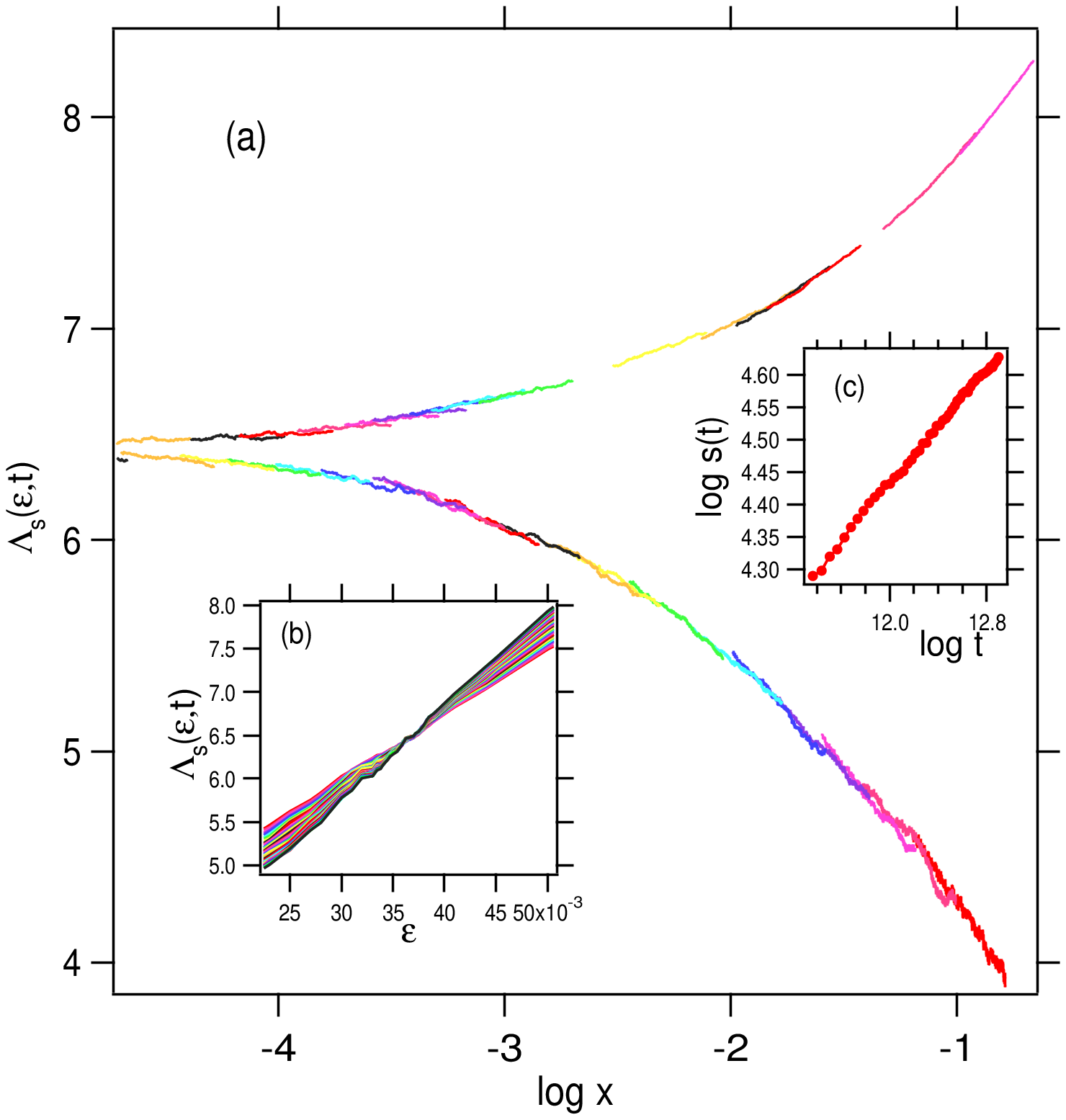}
\caption{(Color online)
The results of the critical scaling analysis for 
trichromatically perturbed AM ($M=3$) with $W=0.5(<W^*)$.
(a)The same scaled MSD $\Lambda_s(\eps,t)$ 
as a function of $x=\xi_0|\eps_c^{AM}-\eps|^{-\nu}t^{\alpha/2\nu}$ for some values 
of $\eps$, where $\xi_0$ is the localization length in the unperturbed case.
(b)The scaled MSD $\Lambda_s(\eps,t)$ with $\alpha=0.51$ 
as a function of $\eps$  for some pick up times.
The crossing point is $\eps_c^{AM} \simeq 0.036$.
(c)$s(t)$ as a function of $t$.
The critical exponent $\nu\simeq 1.18$ is determined least-square fit of (b).
}
\label{fig:c3-w05-nu}
\end{center}
\end{figure}

\begin{figure}[htbp]
\begin{center}
\includegraphics[width=7.0cm]{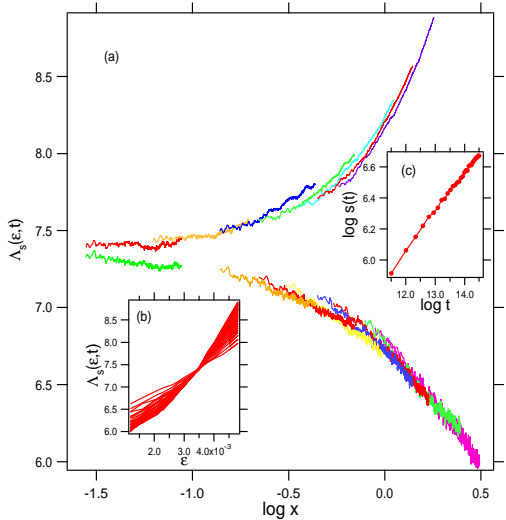}
\caption{(Color online)
The results of the critical scaling analysis for 
polychromatically perturbed AM ($M=7$) with $W=2.0(>W^*)$.
(a)The same scaled MSD $\Lambda_s(\eps,t)$ 
as a function of $x=\xi_0|\eps_c^{AM}-\eps|^{-\nu}t^{\alpha/2\nu}$ for some values 
of $\eps$, where $\xi_0$ is the localization length in the unperturbed case.
(b)The scaled MSD $\Lambda_s(\eps,t)$ with $\alpha=0.25$ 
as a function of $\eps$  for some pick up times.
The crossing point is $\eps_c^{AM} \simeq 0.0034$.
(c)$s(t)$ as a function of $t$.
The critical exponent $\nu\simeq 0.49$ is determined least-square fit of (b).
}
\label{fig:c7-w20-nu}
\end{center}
\end{figure}

Further, Fig.\ref{fig:c3-w05-nu} and \ref{fig:c7-w20-nu} 
displays the results of the finite-time critical scaling analysis for 
trichromatically perturbed AM of  $M=3$ with $W=0.5(<W^*)$
and AM ($M=7$) with $W=2.0(>W^*)$, respectively.
As a result, even in the AM, the OPST is well established for the LDT
regardless of the number of colors $M$ and the disorder strength $W$. 
The localization critical exponent $\nu$ obtained 
is almost similar if $M$ is the same. 
The result strongly suggests that the LDT is a universal transition phenomenon.

\vspace{1cm}


For $M=2$ to $M=7$, in LDT of SM and AM, the critical exponents $\nu$ 
obtained from the critical scaling analysis 
are arranged in table \ref{table:phase1}.
The results of  the critical exponent of the $d-$dimensional Anderson transition 
are also cited from various literatures.
It can be seen that in the perturbed SM and AM with $M$ color modes
the critical localization exponent of LDT   
tend to be similar to the Anderson transition of the $d(=M+1)-$dimensional 
random system.
In other words, the critical exponent $\nu$ of LDT decreases with $M \to \infty$.
In the case of the $d-$dimensional Anderson transition, 
at least in $d \to \infty$, the mean field approximation is in exact, 
and it is considered asymptotic to the result of SCT $\nu=1/2$.
This can also be imagined from the fact that in the Anderson transition, 
spatial connections are important in higher dimensions and the quantum interference 
effect fades.
However, in LDT in SM and AM, the exponent tend to decrease 
to a value smaller than $\nu=1/2$ predicted by SCT as $M$ increases.
Note that in these cases, Harris' inequality $\nu \geq \frac{2}{d}$ is not broken
\cite{harris74}.

\begin{table*}[htbp]
\begin{center}
 \begin{tabular}{lcccccc}
\hline \hline
    & M=2 & M=3 & M=4 & M=5 & M=6 & M=7　\\ \hline 
SM($K=3.1,\hbar=0.24$)  & 1.37 & 0.95 & 0.70  & 0.50 & 0.50 & 0.40\\ 
Ref.\cite{chabe08}   & 1.58 & 1.15 & --  & -- & -- & -- \\  
Ref.\cite{borgonovi97}   & 1.537 & 1.017 & --  & -- & -- & -- \\ 
AM(W=0.5)  & 1.46 & 1.18 & 0.80  & 0.62 &0.53 & 0.41\\ 
AM(W=2.0)  & 1.48 & 1.01 & 0.88  &0.65 & 0.57 & 0.49\\ \hline
    & d=3 & d=4 & d=5 & d=6 & d=7 & d=8　\\ \hline
Ref.\cite{markos06}  & 1.57 & 1.12 & 0.93 & -- & -- & -- \\ 
Ref.\cite{garcia07}   & 1.52 & 1.03 & 0.84 & 0.78 & -- & -- \\ 
Ref.\cite{slevin14}  & 1.57 & 1.15 & 0.97 & -- & -- & -- \\  
Ref.\cite{tarquini17}  & 1.57 & 1.11 & 0.96 & 0.84& -- & -- \\ \hline 
 \hline
 \end{tabular}
 \caption{
The critical exponents numerically obtained by the scaling analysis 
which characterizes the critical dynamics 
in the polychromatically perturbed SM and AM for $M=2 \sim 7$. 
Ref.\cite{chabe08} and Ref.\cite{borgonovi97}  are results 
has already been reported for SM. 
The lower four lines show the critical exponents numerically obtained 
for the $d-$dimensional disordered systems \cite{markos06,garcia07,slevin14,tarquini17}.
}
\label{table:phase1}
\end{center}
 \end{table*} 

\section*{Acknowledgments}
This work is partly supported by Japanese people's tax via JPSJ KAKENHI 15H03701,
and the authors would like to acknowledge them.
They are also very grateful to Dr. T.Tsuji and  Koike memorial
house for using the facilities during this study.




\begin{thebibliography}{00}
\bibitem{footnote-epsc}
The value of  the diffusion index $\alpha$ in the $d-$dimensional Anderson transition
 is not yet known. 
However, it is known that in the Anderson model of the disorder strength $W$, 
the dimensionality of the critical strength $W_c$ is $W_c \sim 4(2d-1)\log(2d-1)$
for large $d$, which corresponds to the behavior in the large-connectivity limit
\cite{garcia07,tarquini17}.
\end{thebibliography}
\end{document}